\pdfoutput = 1
\documentclass[11pt]{article}
\usepackage[margin=1.0in]{geometry}
\usepackage{amsmath,amssymb,graphicx}
\usepackage{nccmath}
\usepackage{hyperref}
\usepackage{graphicx}
\graphicspath{{Figures/}{../Figures/}}

\hypersetup{
  colorlinks   = true, 
  urlcolor     = blue, 
  linkcolor    = blue, 
  citecolor   = blue 
}

\usepackage[T1]{fontenc}

\usepackage[utf8]{inputenc} 
\usepackage{upgreek}
\usepackage{amsmath, amsfonts, mathtools, amssymb}
\usepackage{physics}
\usepackage{slashed}
\usepackage{dsfont}
\usepackage{feynmp-auto}
\usepackage{xcolor}

\usepackage{tikz}
\usetikzlibrary{trees}
\usetikzlibrary{decorations.pathmorphing}
\usetikzlibrary{decorations.markings}
\usetikzlibrary{decorations.pathreplacing}

\usepackage[font=small,labelfont=bf]{caption}
\usepackage{subcaption}
\usepackage{cite}

\usepackage{mdframed}
\usepackage{pdflscape}
\usepackage[section]{placeins}
\usepackage[normalem]{ulem}

\newcommand{\iu}{{\mathrm i}}

\newcommand{\be}{\begin{equation}}
\newcommand{\ee}{\end{equation}}
\def\bge{\begin{equation}}
\def\ede{\end{equation}}
\def\bga{\begin{aligned}}
\def\eda{\end{aligned}}
\def\bgp{\begin{pmatrix}}
\def\edp{\end{pmatrix}}
\def\bgs{\begin{subequations}}
\def\eds{\end{subequations}}

\def\mb{\mathbf}

\def\la{\langle}\def\ra{\rangle}

\def\to{\rightarrow}

\def\al{\alpha}

\def\de{\delta}

\def\si{\sigma}

\usepackage[greek,english]{babel}
\newcommand{\cpi}{\text{\greektext p}}

\newcommand{\Sec}[1]{Sec.~\ref{#1}}

\newcommand{\Fig}[1]{Fig.~\ref{#1}}
\newcommand{\Eq}[1]{Eq.~(\ref{#1})}

\definecolor{goodgreen}{rgb}{0,.6,0.4}

\title{\Large\textbf{Missing Scalars at the Cosmological Collider}
}

\author{Qianshu Lu$^{a}$\footnote{qianshu$\_$lu@g.harvard.edu},~~~ Matthew Reece$^{a}$\footnote{mreece@g.harvard.edu},~~~ Zhong-Zhi Xianyu$^{b}$\footnote{zxianyu@tsinghua.edu.cn}\\[2mm]
$^{a}$\normalsize{\emph{Department of Physics, Harvard University, 17 Oxford Street, Cambridge, MA 02138, USA}}\\[2mm]
$^{b}$\normalsize{\emph{Department of Physics, Tsinghua University, Beijing 100084, China}}
}
\date{ }

\begin{document}
\maketitle

\begin{abstract}
  Light scalar fields typically develop spatially varying backgrounds during inflation. Very often they do not directly affect the density perturbations, but interact with other fields that do leave nontrivial signals in primordial perturbations. In this sense they become ``missing scalars'' at the cosmological collider. We study potentially observable signals of these missing scalars, focusing on a special example where a missing scalar distorts the usual oscillatory features in the squeezed bispectrum. The distortion is also a useful signal distinguishing the de Sitter background induced thermal mass from a constant intrinsic mass.
\end{abstract}

\section{Introduction}
\label{sec:intro}

 It is widely held that a period of exponentially fast cosmic expansion, known as cosmic inflation, occurred prior to the current phase of thermal expansion. The energy scale of  inflation could have been as high as $\order{10^{13}}$ GeV in terms of the Hubble parameter $H$, making inflation physics a truly ``high energy frontier.'' Information about high energy processes during inflation can be carried by the primordial fluctuations. In typical inflation scenarios, the primordial fluctuations were generated from the quantum fluctuations of the spacetime fields. Once produced, they were quickly redshifted to superhorizon scales and remained almost  constant. After the transition to the current thermal big bang phase, the primordial fluctuations re-entered the horizon and sourced the large scale inhomogeneity that we can observe today. 
 
A natural question to ask is how we can make use of the high energy of inflation to probe fundamental physics, especially new physics beyond the Standard Model (SM). It has been known for quite a long time that inflationary expansion can not only generate curvature fluctuations, but also produce on-shell states with masses up to the Hubble scale. Once produced, the heavy particle quickly loses almost all its momentum by cosmic expansion, and its mode function then oscillates with a fixed frequency $m$. If this heavy state is coupled to the curvature fluctuation $\zeta$, it can leave a characteristic oscillatory shape in various soft limits of $n$-point correlators of $\zeta$. Then one may hope to extract the mass and spin and other information about the heavy states from the $n$-point correlators. Recently, there has been a revived interest in studying these objects, with a focus on the oscillatory signal of heavy states. This has been dubbed ``cosmological collider physics,'' since the signal does in some sense resemble a resonant peak of on-shell particle production at a real collider~\cite{Meerburg:2019qqi,Chen:2009we,Chen:2009zp,Chen:2012ge,Pi:2012gf,Gong:2013sma,Arkani-Hamed:2015bza,Chen:2016nrs,Lee:2016vti,Chen:2016uwp,Chen:2016hrz,An:2017hlx,Kumar:2017ecc,Chen:2017ryl,Chen:2018xck,Wu:2018lmx,Li:2019ves,Lu:2019tjj,Hook:2019zxa,Hook:2019vcn,Kumar:2019ebj,Wang:2019gbi,Wang:2019gok,Wang:2020uic,Li:2020xwr,Wang:2020ioa,Fan:2020xgh,Aoki:2020zbj,Bodas:2020yho,Meerburg:2016zdz,Kogai:2020vzz,Arkani-Hamed:2018kmz,Baumann:2019oyu,Baumann:2020dch}. 

In this paper we extend the reach of cosmological collider physics to light bosons that may or may not couple directly to the curvature fluctuation $\zeta$. Scalar fields are ubiquitous in physics beyond the SM, with good theory motivations and interesting phenomenology \cite{Essig:2013lka}. We haven't seen any of them, aside from the SM Higgs boson, because they are either very weakly coupled to the SM or too heavy to be produced in our current experiments. On the other hand, bosons with $m<H$ can be copiously produced by inflationary expansion, irrespective of their SM couplings. With $H$ up to $\order{10^{13}}$ GeV, this greatly enlarges the parameter space of boson production in both the mass and the couplings. 

After their production, sufficiently light scalar fields can survive the inflationary expansion without much dilution. A scalar as light as $\sim0.3H$ can survive 60 e-folds of inflation, and lighter scalars can survive more. But very often, their energy density during inflation is subdominant so that they do not directly affect the density fluctuations. It is also often the case that we can't see them directly today, either because they decay quickly in the thermal big-bang phase or because they are very weakly coupled.\footnote{In the case that they do survive the thermal big-bang and are also abundantly produced, one can look for them in dark matter isocurvature modes \cite{Li:2019ves,Li:2020xwr,Lu:2021gso}.} Even when they are coupled to the inflaton, the inflaton bispectrum mediated by these light scalar fields lacks the oscillatory signature like that mediated by heavier fields. In this sense they become ``missing scalars.'' This terminology is inspired by a direct parallel with a missing energy process in terrestrial collider, as explained later around \Eq{eq:missingScalar}.

In this paper we provide an example of ``missing scalar signal'' at the cosmological collider, by introducing a second and heavier scalar field. The heavier scalar $\si$ couples directly to the inflaton as in the original quasi-single-field inflation~\cite{Chen:2009we,Chen:2009zp}, so that we can see corresponding oscillation signals of $\si$ in inflaton correlators. We then assume that $\si$ interacts with a light scalar $\chi$, our missing scalar, via a dimension-4 coupling $g\si^2\chi^2$. The light scalar $\chi$ then gives a space-dependent mass correction to $\si$, and it is this space-dependent mass that will leave a distinct signal in the inflaton correlator. The heavy field $\si$ thus serves as a probe of $\chi$ that bridges the missing scalar with the curvature perturbation. We note that the use of a scalar field $\si$ is inessential; similar signals should also appear for heavy particles with nonzero spin. We leave this more general case for future studies.

There is a separate motivation for our study of the signal described above. Common lore of cosmological collider physics holds that measuring the frequency $\nu$ of the oscillatory signal tells the mass $m$ of the intermediate particle, since there is a one-to-one relation between them for fixed species, e.g., $\nu=\sqrt{(m/H)^2-9/4}$. But the mass inferred in this way is usually ``dressed'' by all kinds of inflationary corrections \cite{Chen:2016hrz}. The most common correction is a type of infrared enhanced loop correction from light degrees of freedom, very similar but not identical to the thermal mass in a finite temperature system in flat space \cite{Starobinsky:1994bd,Rajaraman:2010xd,Gorbenko:2019rza,Mirbabayi:2019qtx,Baumgart:2019clc}. This correction is closely related to the Hawking temperature $T=H/(2\pi)$ felt by any inertial observer in dS. For this reason we will call it a dS thermal mass in this paper. In our example $g\si^2\chi^2$, the light field $\chi$ will introduce a mass correction to $\si$ roughly given by $\Delta m_\si^2\sim g H^4/m_\chi^2$. So the question is whether we can separate this frequently-appearing thermal mass correction from the intrinsic mass of $\si$. We will show that this is possible, since the dS thermal mass correction is often associated with a space-dependent mass correction at next order in $g$ which generates a ``missing scalar'' signal. Therefore, by searching for this missing scalar signal, one can try to distinguish a thermal contribution from the total mass.

In the rest of this section, we discuss possible ways to identify a missing scalar $\chi$ from inflaton correlators. We will review a few possible processes and explain why we choose to focus on the 3-point correlator mediated by $\si$. 

Perhaps the simplest process involving a missing scalar $\chi$ would be the following one,
\bge
 \parbox{0.35\textwidth}{\includegraphics[width=0.35\textwidth]{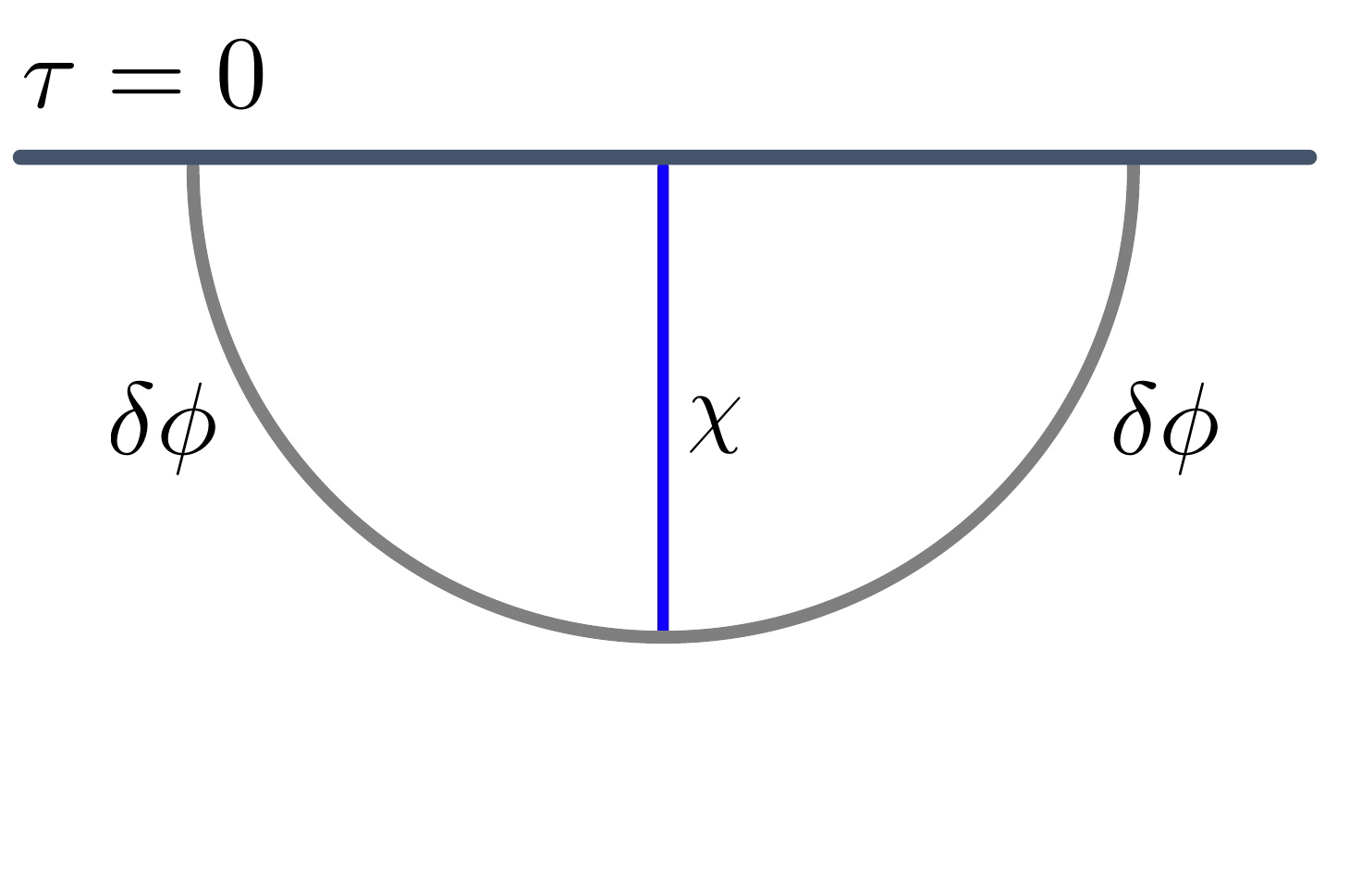}\vspace{-8mm}}\label{eq:missingScalar}
\ede
This is completely in parallel with a missing energy process in a real collider. Here we are assuming that the missing scalar $\chi$ couples to the inflaton fluctuations through a cubic interaction term such as $(\de\phi')^2\chi$. In such a process, the missing scalar $\chi$ could carry away a nonzero 3-momentum, making the effective 2-point correlation of $\de\phi$ momentum non-conserving. Naively this would suggest us to look for a momentum non-conserving component in the power spectrum. But this is never observed, not because such a component is small and below our sensitivity, but because normally we don't have the access to each individual process like this. Instead, we observe statistical average of many such processes. Although the individual process produces a momentum non-conserving 2-point function for $\de\phi$, the average over many of them is isotropic. This also applies to missing energies in real collider events. 

Therefore, we should look for a signal of a missing scalar that survives the statistical average. Given that $\la\chi\ra=0$ up to cosmic variance, the only processes surviving the statistical average are those with no $\chi$ lines propagating to $\tau=0$. At the 2-point level, there are two processes we may want to look at:
\bge
  \parbox{0.35\textwidth}{\includegraphics[width=0.35\textwidth]{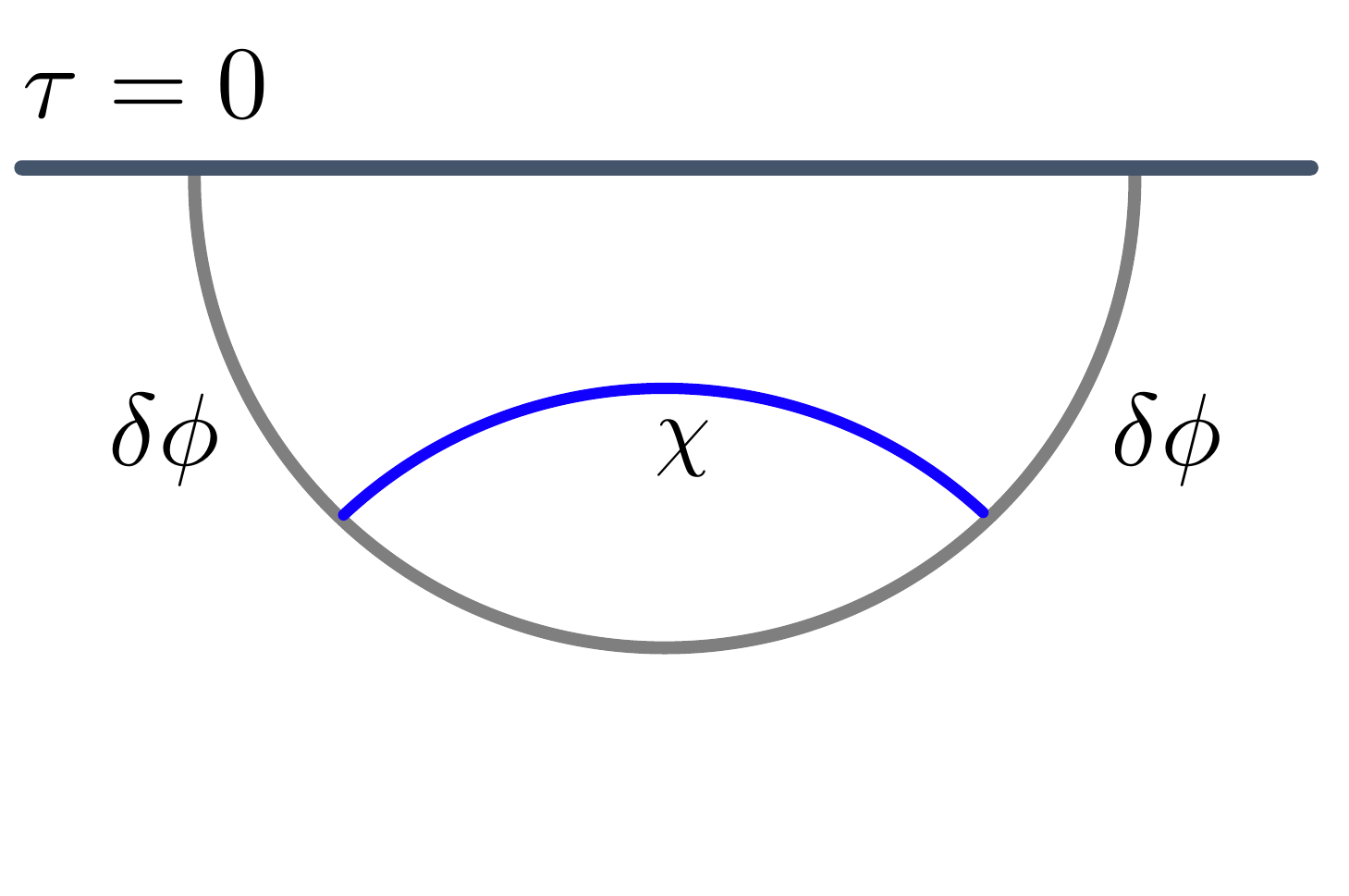}}~~~~~~
  \parbox{0.35\textwidth}{\includegraphics[width=0.35\textwidth]{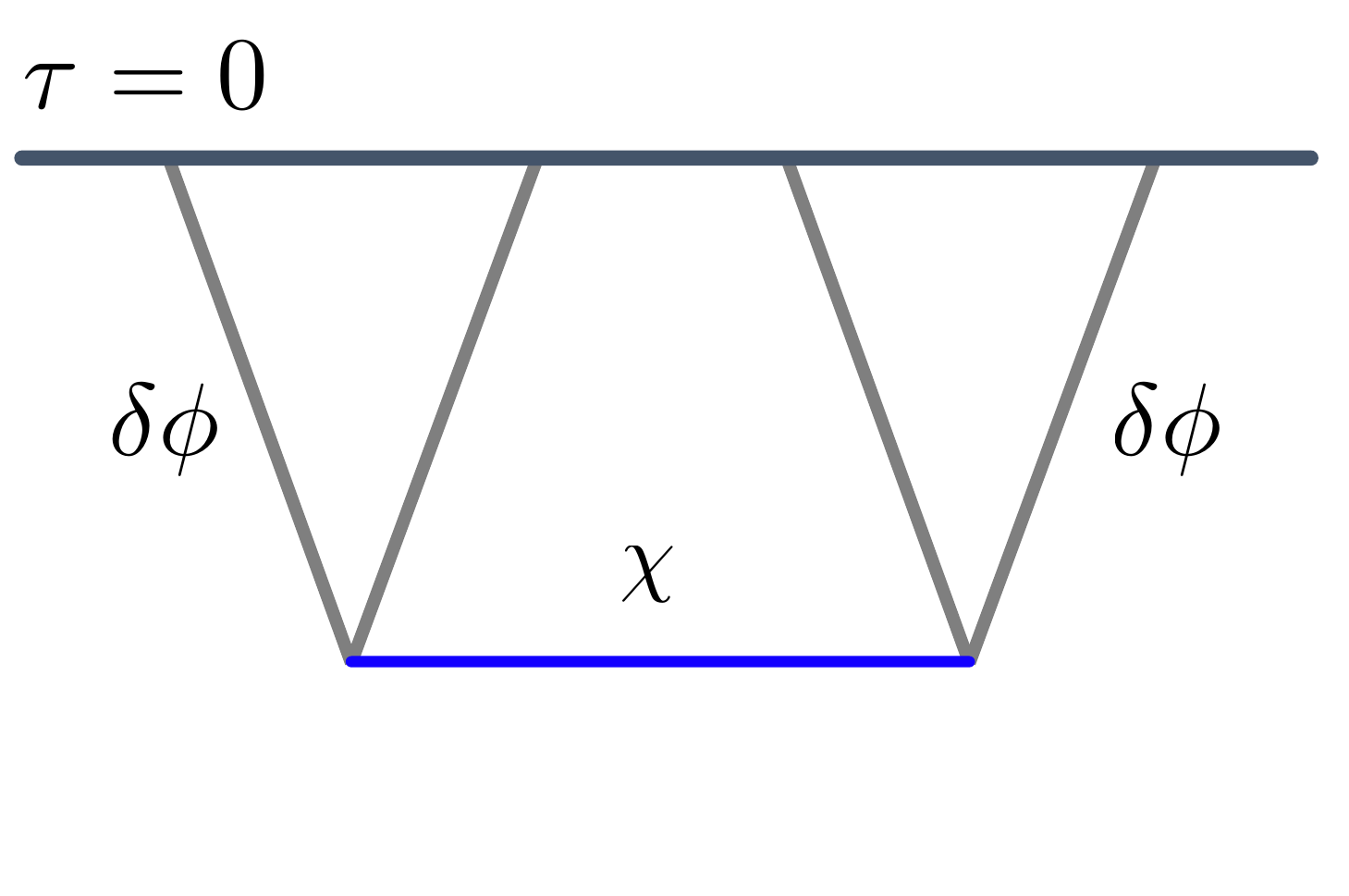}}\vspace{-5mm}
\ede
The left one is a loop correction that is most likely scale invariant up to slow-roll corrections when $\chi$ is light. (The mass of $\chi$ here is an effective slow-roll parameter.) So this correction can be easily mimicked by changing the inflation potential a little bit. Thus, we will consider it a non-observable effect. The right one is nevertheless observable and was studied previously \cite{Jeong:2012df,Dai:2013ikl}. It essentially says that if we measure the 2-point function of $\de\phi$ in two different sky patches then we will get different answers, and the difference can be attributed to the long-distance modulation of $\chi$. But this is in fact a 4-point observable, as is obvious from the graph above. So it is not as easy to measure as 2-point or 3-point correlations. 

One point we want to make is that we could already see interesting effects of missing scalars making use of the ``cosmological collider observables'' at the 3-point level. Maybe the first thing coming to our mind at this point is the following graph,
\bge
  \parbox{0.35\textwidth}{\includegraphics[width=0.35\textwidth]{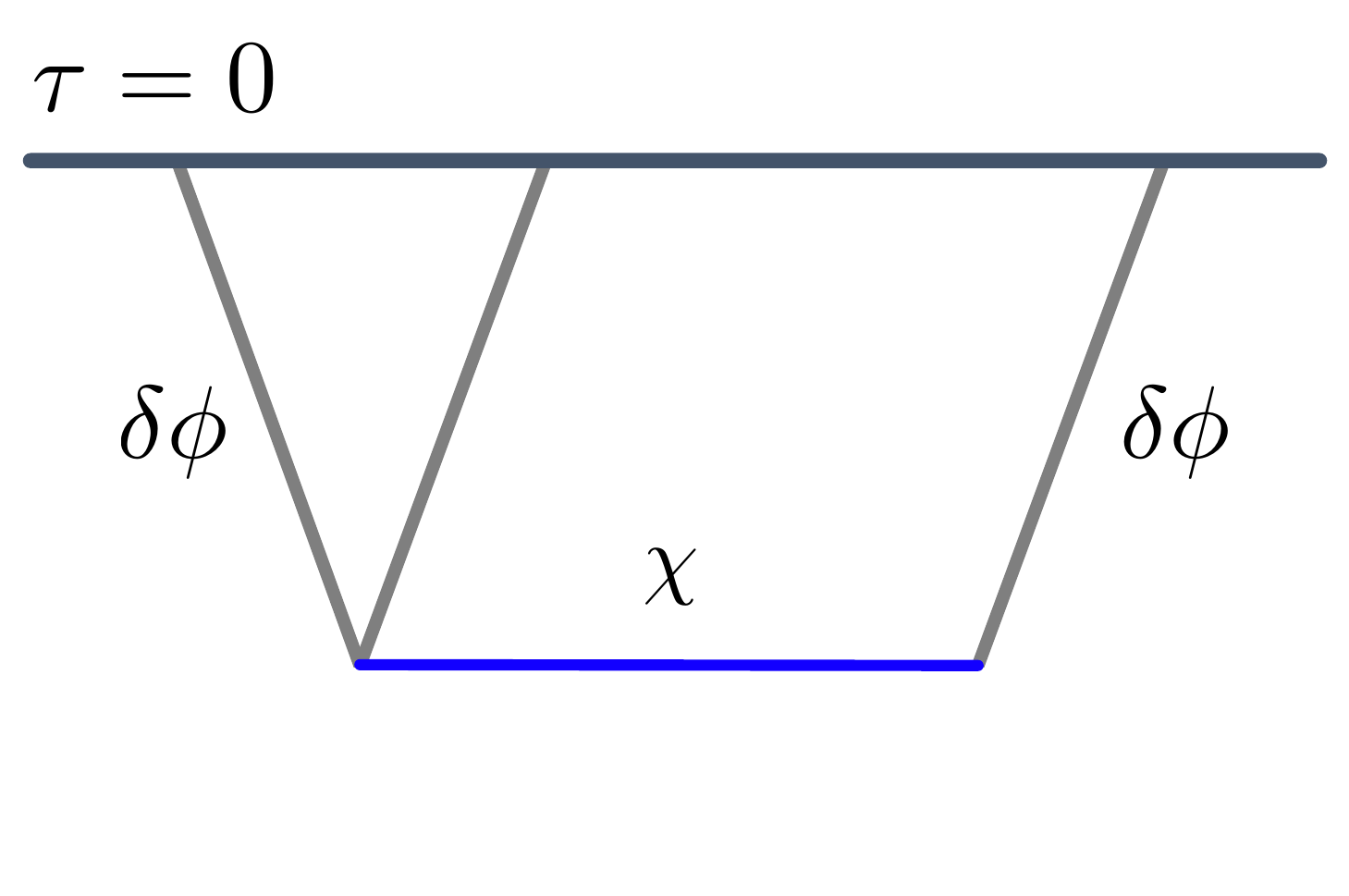}}\vspace{-5mm}\label{eq:tree_three_point}
\ede
This is exactly the quasi-single-field process extensively studied previously \cite{Chen:2009we, Chen:2009zp}. When $\chi$ is light, the squeezed limit of this process will produce a non-analytical scaling in the ratio of soft and hard external momenta. While this is certainly an interesting situation to look into, it nevertheless requires a direct coupling between the light field $\chi$ and the inflaton $\de\phi$. Below we will turn to other interesting possibilities at 3-point level which do not require a direct coupling between $\chi$ and $\de\phi$. It would be interesting to further study the case in which both of these signals could be detected.

The starting point of our investigation is an additional field $\si$ with $m_\si\sim H$, which could have nonzero spin, but which we will take to be a scalar for simplicity. Then we assume that $\si$ get mass from $\chi$. The effective mass of $\si$ is then $m_\si^2(\chi)=m_{\si0}^2+g\la\chi^2\ra$. Here $m_{\si0}$ is a constant mass independent of $\chi$. We introduce this piece only to ensure that $\si$ fluctuations do not strongly backreact on the $\chi$ potential.

For a light scalar $\chi$ we would expect $\la\chi^2\ra\sim H^2$ and for $g\sim\order{1}$ we do get an $\order{H^2}$ contribution to $m_\si^2$. But this is quite different from the mass from a classical vev $g\la\chi\ra^2$. In the latter case we do have nonzero $\chi$ at each space point, and if the potential at the $\chi$ vev is steep enough, this will give a constant shift of $m_\si$. On the contrary, the case we are considering actually corresponds to $\la\chi\ra=0$ but $\la\chi^2\ra\neq 0$, and there will be spatial variations in $\la\chi^2\ra$. Telling these two different mass corrections apart is therefore another separate motivation for our study which may have interesting phenomenological consequences (such as if SM Higgs would develop a second vev at high energies \cite{Hook:2019zxa,Hook:2019vcn}).

Now since we are interested in the spatial variation of $\si$ mass, we may try to use the cosmological collider to measure this mass for different patches. But again, comparing masses over different patches is actually measuring a 6-point correlation, as shown below.
\bge
  \parbox{0.35\textwidth}{\includegraphics[width=0.35\textwidth]{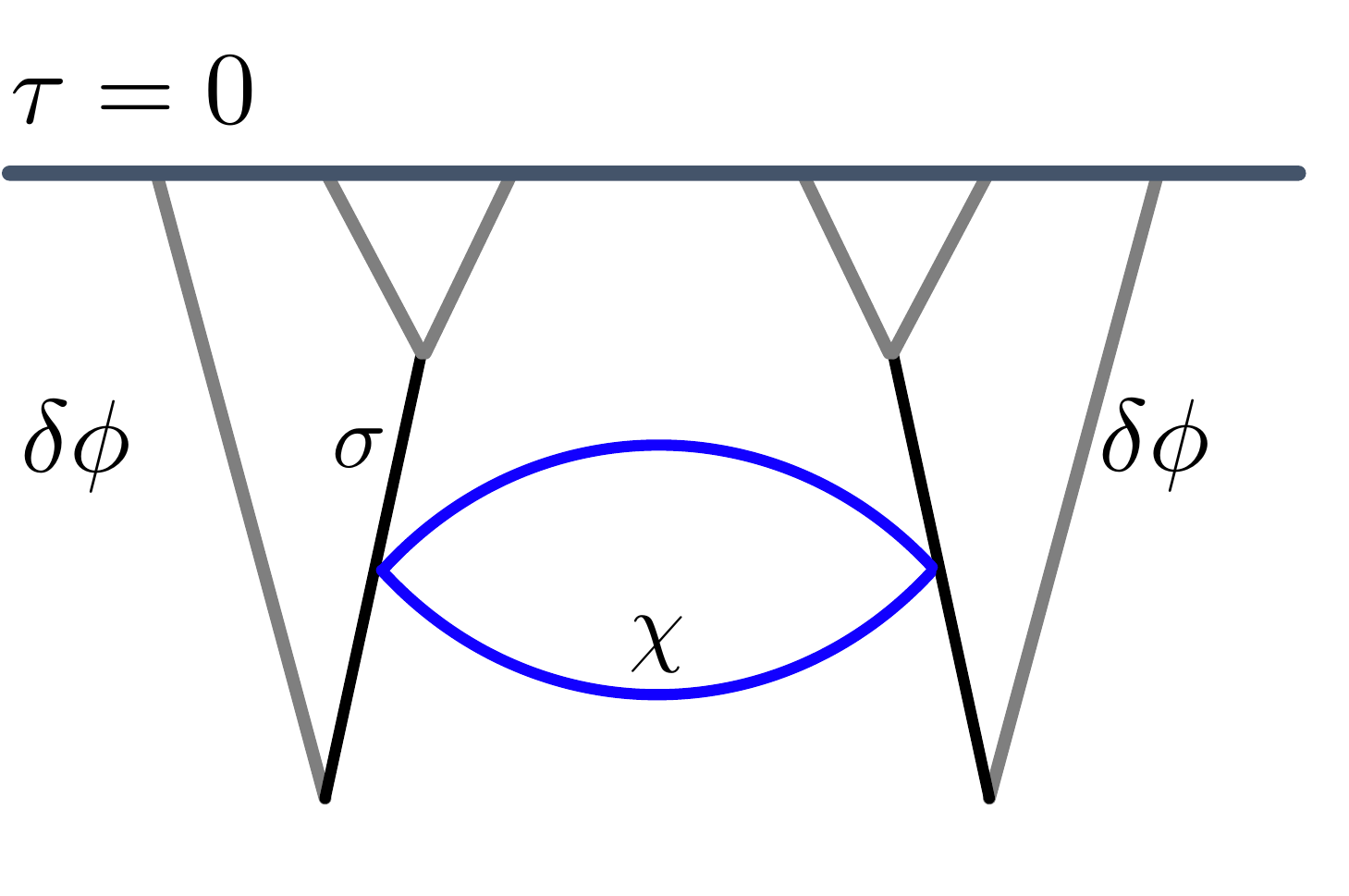}}
\ede
Instead, we want to stay at the 3-point level. At this level the simplest process that can tell a true mass from a spatially varying mass would be the following one. 
\bge
  \parbox{0.35\textwidth}{\includegraphics[width=0.35\textwidth]{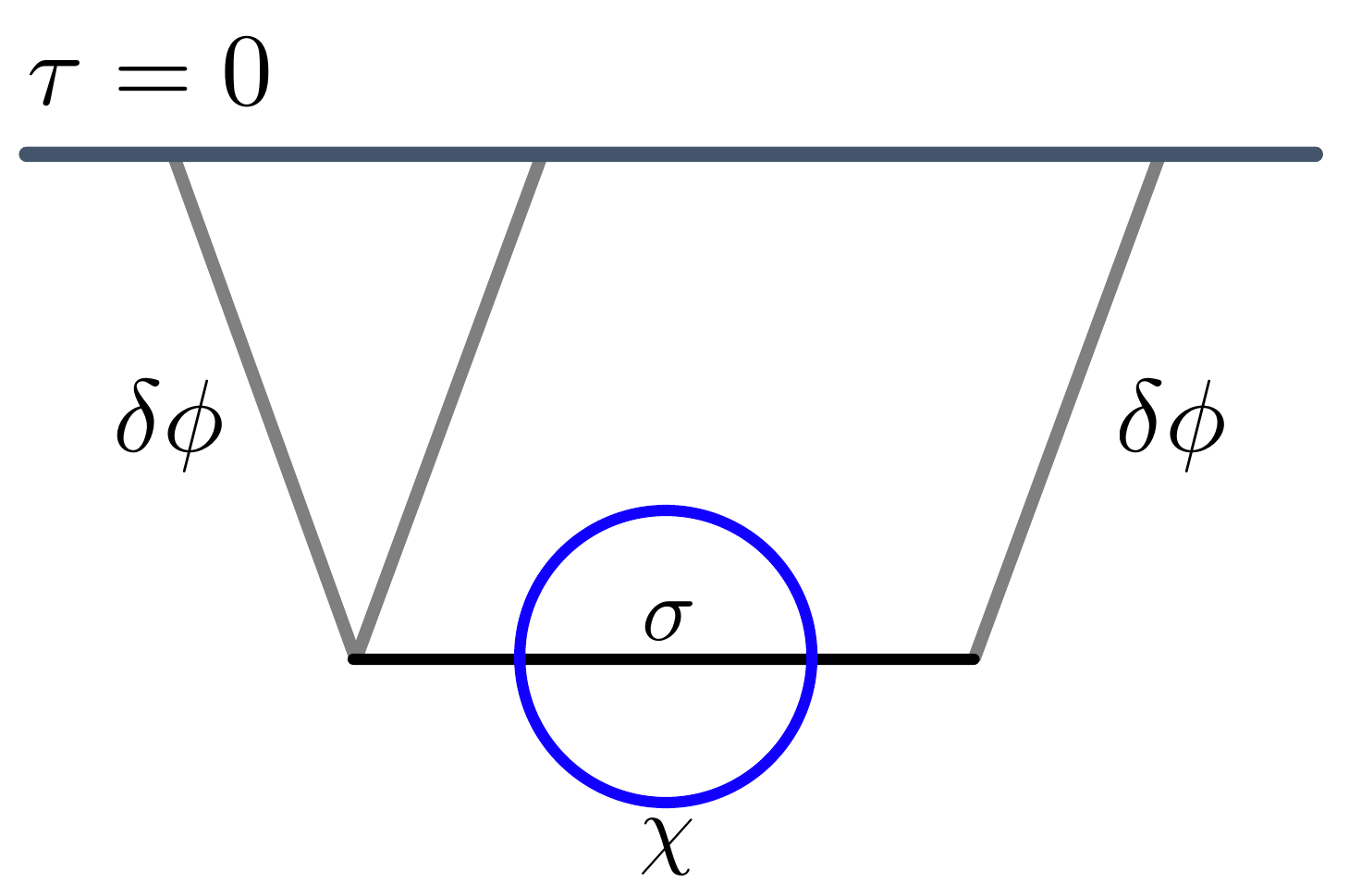}}\label{eq:bispec}
\ede
We will thus study this process in detail in the rest of this paper, and we will show that this process produces a distinct cosmological collider signal in the squeezed limit $k_1=k_2\gg k_3$ of the bispectrum. The signal can be expressed in terms of the dimensionless shape function $S$ of the 3-point correlator as\footnote{The shape function $S$ we are using in this paper is related to the correlator of curvature fluctuations by 
\bge
  \la\zeta_{\mb k_1}\zeta_{\mb k_2}\zeta_{\mb k_3}\ra=(2\pi)^3\de^{(3)}(\mb k_1+\mb k_2+\mb k_3)\frac{(2\pi)^4 P_\zeta^2}{k_1^2k_2^2k_3^2}S(k_1,k_2,k_3),
\ede 
where $P_\zeta\simeq 2\times 10^{-9}$ is the size of the nearly scale invariant power spectrum. }
\be \label{eq:result}
S_{\text{signal}}(k_1=k_2 \gg k_3) = \left(\frac{k_3}{k_1}\right)^{\frac12+\iu\nu_{\sigma}+\alpha},
\ee 
where $\nu_\si=\sqrt{m_\si^2/H^2-9/4}$ is the well-known oscillation frequency related to the (dressed) mass $m_\si$ of the $\si$ field. Our signal features a positive exponent $\al>0$, which is absent in the ordinary cosmological collider signal induced by constant masses.\footnote{Similar behavior in the squeezed bispectrum with $\al\neq 0$ also appears in the non-standard clock signal of alternatives to inflation models \cite{Wang:2020aqc}. But there the signal is typically accompanied by explicit scale dependence, while our signal is manifestly scale invariant up to slow-roll corrections. } This $\al$ dependence signals the loss of correlation at superhorizon scales. As we will show below, this is essentially due to the space-dependent mass correction generated by a fluctuating light field $\chi$.

The rest of the paper is organized as follows: In \Sec{sec:model}, we explain the model that we study and the general structure of perturbative calculations in the model. In \Sec{sec:correction}, we compute the one-loop correction to the heavy scalar propagator and explain how it is altered by fluctuations of the light scalar. In \Sec{sec:observability}, we provide a preliminary estimate of the observability of the signal. Finally, in \Sec{sec:conclusions} we offer concluding remarks.

\section{The Model and Setup}
\label{sec:model}

Consider a simple model with two massive scalars, $\sigma$ and $\chi$, with a quartic coupling between them. We work in 4d Euclidean de Sitter spacetime, because the Euclidean de Sitter continuation makes the physics of certain infrared divergences more tractable~\cite{Rajaraman:2010xd, Marolf:2010zp}. In particular, we will find the decomposition of the field into its Euclidean zero mode and non-zero modes useful below.
The metric is 
\be
\dd s^2 = \frac{1}{H^2}\left[\dd \tau^2 + (\sin \tau)^2 \dd \Omega_3^2\right],
\ee
where $H$ is the Hubble constant. Because the $D$-dimensional Euclidean de Sitter spacetime is a $D$-dimensional sphere, the field momentum is quantized. The free two-point function of a field $f$ of mass $m_{f,0}$ can thus be written as a sum over the discrete momenta as
\be
\begin{split}
\expval{f(x)f(y)} \equiv \Delta^f(x, y)&= H^2\sum_{\vec{L}}\frac{ Y_{\vec{L}}(x)Y_{\vec{L}}^*(y)}{m_{f,0}^2/H^2 + L(L+3)}\\
&= H^2 \frac{\Gamma\left(\frac{3}{2}\right)}{2\cpi^{\frac{5}{2}}}\sum_{L=0}^{\infty}\frac{L+\frac{3}{2}}{m_{f,0}^2/H^2+L(L+3)}C_L^{\frac32}(Z_{xy}),
\end{split}
\ee
where $\vec{L} = (L_4, L_3, L_2, L_1)$ is the set of four discrete momenta with $L_4\leq L_3 \leq L_2 \leq |L_1|$, and $L=L_4$ is the total angular momentum. $Y_{\vec{L}}(x)$ is the equivalent of spherical harmonics on this 4-dimensional sphere. $C_L^{d/2}$ is a Gegenbauer function for general complex index $L$, and $Z_{xy}$ is the embedding distance between $x$ and $y$, defined by
\be
Z_{xy}\equiv \cos\tau_x \cos\tau_y+\sin\tau_x\sin\tau_y(\vec{x}\cdot \vec{y}).
\ee
The value of $m_{f, 0}^2$ determines the large distance behavior of $\Delta^f(x,y)$,
\be
\Delta^f(x, y)\xrightarrow{|Z_{xy}|\rightarrow \infty}\begin{cases}
|Z_{x,y}|^{-\frac{3}{2}\pm \iu \sqrt{\frac{m^2_{f,0}}{H^2}-\frac{9}{4}}} &m^2_{f, 0}> \frac{9}{4}H^2\\
|Z_{x,y}|^{-\frac{3}{2}+ \sqrt{\frac{9}{4}-\frac{m^2_{f,0}}{H^2}}} &m^2_{f, 0}< \frac{9}{4}H^2.\label{eq:scaling}
\end{cases}
\ee

\begin{figure}
\centering
\includegraphics[width=\textwidth]{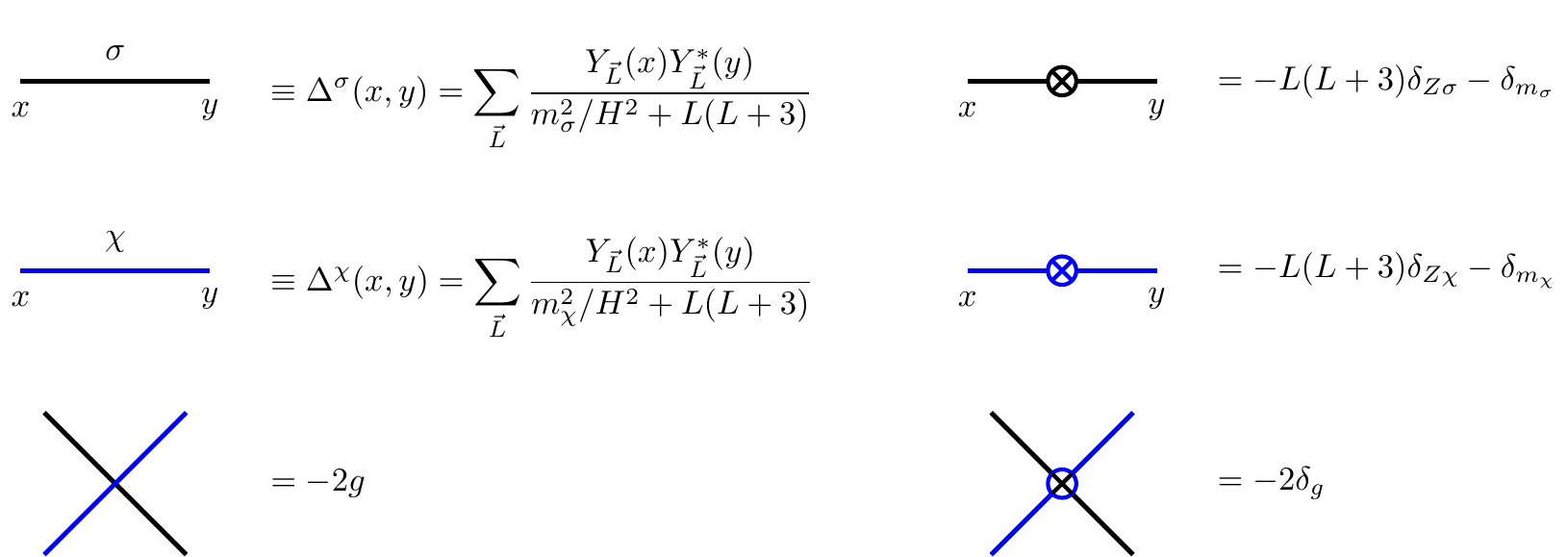}
\caption{Feynman rules from the Lagrangian in \Eq{eq:lag}.}
\label{fig:rules}
\end{figure}

We will be discussing various loop corrections to the two-point functions of $\sigma$ and $\chi$, where ultraviolet divergences will arise. To make explicit what quantities are physical, we will work in the framework of renormalized perturbation theory, with the following Lagrangian
\be
\begin{split}
\mathcal{L} =& g_{\mu\nu}\frac{1}{2}\partial^{\mu}\sigma\partial^{\nu}\sigma +g_{\mu\nu}\frac{1}{2}\partial^{\mu}\chi\partial^{\nu}\chi + \frac{1}{2}m_{\sigma}^2\sigma^2 + \frac{1}{2}m_{\chi}^2\chi^2 + \frac{g}{2}\sigma^2\chi^2\\
+&g_{\mu\nu}\frac{1}{2}\delta_{Z\sigma}\partial^{\mu}\sigma\partial^{\mu}\sigma+g_{\mu\nu}\frac{1}{2}\delta_{Z\chi}\partial^{\mu}\chi\partial^{\mu}\chi+\frac{1}{2}\delta_{m_\sigma}\sigma^2 + \frac{1}{2}\delta_{m_\chi}\chi^2+\frac{\delta_g}{2} \sigma^2\chi^2.
\end{split}\label{eq:lag}
\ee
The parameters $m_{\sigma}$, $m_{\chi}$ are defined to be the \textit{physically measured} mass. This means the \textit{loop-corrected} two point function has the following large distance scaling
\be
\Delta^f(x, y)_{\text{loop-corrected}}\xrightarrow{|Z_{xy}|\rightarrow \infty}|Z_{xy}|^{\alpha},\quad \begin{cases}
\Im(\alpha) = \pm\sqrt{\frac{m_f^2}{H^2}-\frac{9}{4}} & m_f^2 > \frac{9}{4}H^2\\
\Re(\alpha) = -\frac{3}{2}+\sqrt{\frac{9}{4}-\frac{m_f^2}{H^2}} & m_f^2 < \frac{9}{4}H^2.\\
\end{cases}
\ee
This way of defining the mass renormalization condition is natural if we recall that in Minkowski space, the mass of a field also determines the rate of exponential decay of two-point correlation functions at large distances. The parameters on the second line of \Eq{eq:lag} are counterterms to enforce the above renormalization conditions for $m_{\sigma}$ and $m_{\chi}$ at every order in perturbation theory. The counterterms contribute additional Feynman rules on top of the familiar ones from the physical parameters, as shown in \Fig{fig:rules}. The interaction strength $g$ will also be renormalized by loop corrections, but it does not suffer from power corrections so the effects are mild, and we will not comment any further on this.

In addition to the coupling with $\chi$, we assume that $\sigma$ has the necessary coupling with the inflaton in order to mediate an inflaton three-point function through a diagram similar to that shown in \Eq{eq:tree_three_point}. The inflaton bispectrum mediated by a freely propagating $\sigma$ is well-known \cite{Chen:2009we, Chen:2009zp, Arkani-Hamed:2015bza}
\be
S(k_1=k_2 \gg k_3) \propto \left(\frac{k_3}{k_1}\right)^{\frac{1}{2}\pm\mu_{\sigma}},\quad \mu_{\sigma} = \sqrt{\frac{9}{4}-m_{\sigma}^2}.
\ee
We fix $m_{\sigma}^2>9/4H^2$ so $\mu_{\sigma} = \iu \nu_{\sigma} = \iu \sqrt{m_{\sigma}^2/H^2-9/4}$ is imaginary, and the inflaton bispectrum mediated by $\sigma$ is going to have an oscillatory shape in the squeezed limit. The magnitude of the inflaton bispectrum will be a free parameter in our model since we do not specify the coupling strength between $\sigma$ and the inflaton. We are looking for a distinctive modification of the shape function due to the interaction of $\sigma$ with $\chi$.

\begin{figure}
\centering
\includegraphics[width=\textwidth]{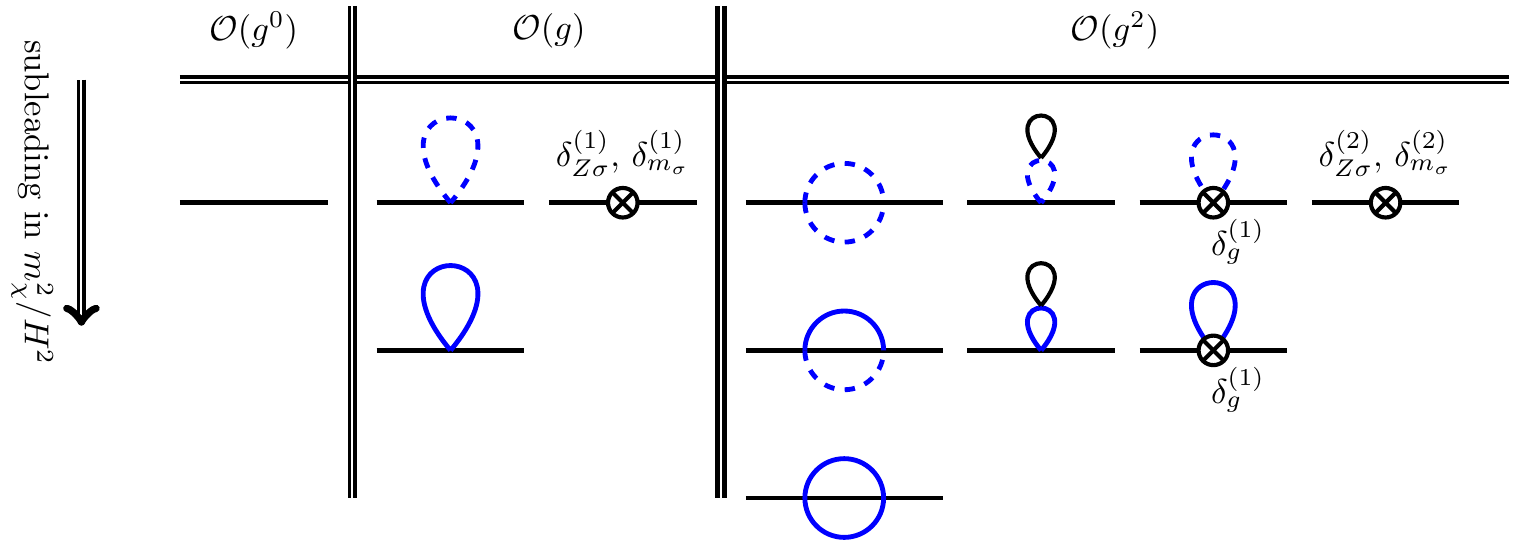}
\caption{Organization of Feynman diagrams in a double perturbation series in $g$ and $m_{\chi}^2/H^2$. Black lines are $\sigma$ and blue lines are $\chi$. For the $\chi$ propagator, we distinguish the $L=0$ mode (dashed line) and the $L\neq 0$ modes (solid line). Diagrams with a $\chi$ zero mode are parametrically larger than the same diagram with the $\chi$ zero mode replaced by a nonzero mode. The counterterm $\delta_g^{(1)}$ is determined by calculation of the $\mathcal{O}(g)$ one-loop correction to the $\sigma^2\chi^2$ interaction, whose diagram is not shown here.}\label{fig:diagrams}
\end{figure}

We will also require $m_{\chi}^2 < H^2$ for the following reason. If we look at the form of the two-point function in \Fig{fig:rules}, we see that when $m_{\chi}^2 < H^2$, the zero mode ($L = 0$) is parametrically enhanced compared to the non-zero ($L \neq 0$) terms:
\be
\frac{1}{m_{\chi}^2/H^2} > \frac{1}{L(L+3)+m_{\chi}^2/H^2}.
\ee
As a consequence, we can organize Feynman diagrams in a double perturbation series, in powers of $g$ and $m_{\chi}^2/H^2$. For example, diagrams that contribute to correction of the $\sigma$ two-point function up to $\mathcal{O}(g^2)$ are shown in \Fig{fig:diagrams}. Going from left to right is the usual perturbative expansion in powers of $g$. At every order $g^n$, $n\geq 1$, all counterterms are fixed to cancel the ultraviolet divergences in the loop diagrams such that the renormalization conditions are met. The order in $g$ for the counterterms is denoted by the superscript $(n)$. At each order in $g$, going from top to bottom is the second perturbative expansion in powers of $m_{\chi}^2/H^2$, where diagrams with more zero-mode $\chi$ propagators are enhanced compared to diagrams with the same topology but fewer zero-mode $\chi$ propagators. This is computationally advantageous, because the zero-mode $\chi$ propagator is a constant,
\be
\expval{\chi(x)\chi(y)}_{\text{zero mode}} \equiv \Delta^{\chi}_0 = \frac{3}{8\cpi^2}\frac{1}{m_{\chi}^2/H^2},
\ee
which means diagrams with more zero-mode $\chi$ propagators are easier to calculate. 

\section{Local and non-local correction to two-point functions}
\label{sec:correction}

The Feynman diagrams listed in \Fig{fig:diagrams} can be separated into two categories depending on the type of corrections they give to $\expval{\sigma(x)\sigma(y)}$. The first type is a local correction that manifest as a constant shift in $\sigma$ mass. Diagrams with only one interaction point with the external $\sigma$ propagator belong to this category, which are everything except the sunset type diagrams on the first column of the $\mathcal{O}(g^2)$ block. For example, consider the $\mathcal{O}(g)$ seagull diagram with a zero mode $\chi$ loop. Using the orthogonality of the 4-dimensional spherical harmonics,
\be
\int \dd x\sqrt{g} Y_{\vec{L}}(x)Y^*_{\vec{M}}(x) =H^{-4}\delta_{\vec{L}\vec{M}},
\ee
the contribution from this diagram is easily calculated:
\be
\begin{split}
\begin{tikzpicture}[line width=1.5pt]
\draw (0.5, 0)--(2.5, 0);
\draw[blue, dashed] (1.5,0)  to [in=50,out=130,loop, distance=1.5cm] (1.5,0);
\end{tikzpicture} &=\expval{\sigma(x)\sigma(y)}_{\text{zero mode seagull}}\\
&=-g\Delta^{\chi}_0\int_z \Delta^{\sigma}(x, z)\Delta^{\sigma}(z, y)\\
&=-g\Delta^{\chi}_0  H^2 \frac{\Gamma\left(\frac{3}{2}\right)}{2\cpi^{\frac{5}{2}}}\sum_{L=0}^{\infty}\frac{L+\frac{3}{2}}{\left(m_{\sigma}^2/H^2+L(L+3)\right)^2}C_L^{\frac32}(Z_{xy}).
\end{split}
\ee
It's clear why this diagram is a mass correction if we look at the 1PI sum of this seagull diagram
\begin{fleqn}

\begin{tikzpicture}[line width=1.5pt]
\draw (-2, 0)--(0, 0);
\draw (0.5, 0)--(2.5, 0);
\draw[blue, dashed] (1.5,0)  to [in=50,out=130,loop, distance=1.5cm] (1.5,0);
\node at (0.25, 0) {$+$};
\node at (2.75, 0) {$+$};
\draw (3, 0)--(6, 0);
\draw[blue, dashed] (4,0)  to [in=50,out=130,loop, distance=1.5cm] (4,0);
\draw[blue, dashed] (5,0)  to [in=50,out=130,loop, distance=1.5cm] (5,0);
\node at (6.5, 0) {$+\cdots$};
\end{tikzpicture}
{\small
\begin{gather}
= H^2 \frac{\Gamma\left(\frac{3}{2}\right)}{2\cpi^{\frac{5}{2}}}\sum_{L=0}^{\infty}\frac{L+\frac{3}{2}}{m_{\sigma}^2/H^2+L(L+3)}C_L^{\frac32}(Z_{xy})\!\left[1-\frac{g\Delta^{\chi}_0 }{m_{\sigma}^2/H^2+L(L+3)}+\left(\frac{g\Delta^{\chi}_0 }{m_{\sigma}^2/H^2+L(L+3)}\right)^2\! \! \!+\cdots\right]\\
= H^2 \frac{\Gamma\left(\frac{3}{2}\right)}{2\cpi^{\frac{5}{2}}}\sum_{L=0}^{\infty}\frac{L+\frac{3}{2}}{m_{\sigma}^2/H^2+L(L+3)+g\Delta_0^{\chi}}C_L^{\frac32}(Z_{xy})\nonumber,
\end{gather}}
\end{fleqn}
where it's manifest that the seagull diagram contributes $g\Delta_0^{\chi}$ to the $\sigma$ mass. To satisfy our renormalization condition that $m_{\sigma}^2$ is the physically observed mass, this means that the counterterm $\delta_{m_{\sigma}}$ must be $-g\Delta^0_{\chi}H^2$ to cancel out the correction from the zero-mode seagull diagram. The seagull diagram with a non-zero mode $\chi$ loop turns out to give a UV divergent correction to $m_{\sigma}^2$, which should similarly be cancelled out by the $\delta_{m_{\sigma}}$ counterterm at $\mathcal{O}(g)$. The same procedure follows for local contributions at $\mathcal{O}(g^2)$. Observationally, we will not be able to tell if the $\sigma$ mass is being corrected by the presence of $\chi$; therefore, the local corrections are not useful for the detection of $\chi$.

Notice that the two-point function of $\chi$ also receives the exact same kind of local correction from its interaction with $\sigma$. The only difference is that separation of the zero-mode and non-zero modes of $\sigma$ is not useful, since $m_{\sigma}^2 > H^2$. The calculation is still simple because the two-point function at coincident points in Euclidean de Sitter space is independent of position (although it is UV divergent), 
\be
\Delta^{\sigma}(x, x)\equiv [\Delta^{\sigma}_{=}].
\ee
Therefore the diagram evaluates to
\be
\begin{split}
\begin{tikzpicture}[line width=1.5pt]
\draw[blue] (0.5, 0)--(2.5, 0);
\draw (1.5,0)  to [in=50,out=130,loop, distance=1.5cm] (1.5,0);
\end{tikzpicture} &=\expval{\chi(x)\chi(y)}_{\text{seagull}}\\
&=-g[\Delta^{\sigma}_{=}]\int_z \Delta^{\chi}(x, z)\Delta^{\chi}(z, y)\\
&=-g[\Delta^{\sigma}_{=}]  H^2 \frac{\Gamma\left(\frac{3}{2}\right)}{2\cpi^{\frac{5}{2}}}\sum_{L=0}^{\infty}\frac{L+\frac{3}{2}}{\left(m_{\chi}^2/H^2+L(L+3)\right)^2}C_L^{\frac32}(Z_{xy}).
\end{split}
\ee
The 1PI sum of this seagull diagram generates a correction to the $\chi$ mass by $m_{\chi}^2\rightarrow m_{\chi}^2 + g[\Delta^{\sigma}_{=}]$. This constant and infinite correction is then cancelled by the counterterm $\delta_{m_{\chi}}$ to satisfy our renormalization condition for $m_{\chi}$. The procedure ensures that the separation of the zero mode and non-zero mode propagators of $\chi$ is meaningful even after the loop corrections are taken into account, since the physical mass of $\chi$ is kept at $m_{\chi}^2<H^2$.

The second type of correction to $\expval{\sigma(x)\sigma(y)}$ is a non-local correction that cannot be mimicked by a constant shift in $m_{\sigma}$, and is the smoking gun signal for $\chi$ that we seek. The leading diagram that contributes such a non-local correction is the sunset type diagram at $\mathcal{O}(g^2)$. Among the three sunset type diagrams, the diagram with two zero-mode $\chi$ propagators is the leading order in the $H^2/m_{\chi}^2$ expansion. However, because the zero-mode $\chi$ propagator is a constant, this diagram is also a local correction which contributes a constant shift in $m_{\sigma}$ that would be removed by the $\delta_{m_{\sigma}}$ counterterm. The leading order diagram for our non-local correction signal is thus the ``half-sunset'' diagram that has one zero-mode $\chi$ propagator.

Since the zero mode propagator is a constant, the “half-sunset” diagram, where we have one $\chi$ zero mode and one $\chi$ nonzero mode, reduces to a constant multiplied by the bubble diagram from a theory with a cubic interaction.
\be
\begin{tikzpicture}[line width=1.5pt]
\draw (0, 0)--(4, 0);
\draw[blue] (1.2,0) arc (180:0:0.8);
\draw[blue, dashed] (1.2,0) arc (180:360:0.8);
\node at (4.5, 0) {\Large$=$};
\draw[blue, dashed] (5, 0)--(8, 0);
\node at (8.5, 0) {\Large$\times$};
\draw (9, 0)--(10, 0);
\draw[blue] (10,0) arc (180:0:0.8);
\draw (10,0) arc (180:360:0.8);
\draw (11.6, 0)--(12.6, 0);
\draw[blue] (1.2,0) arc (180:0:0.8);
\draw[blue, dashed] (1.2,0) arc (180:360:0.8);
\end{tikzpicture}
\ee
Such a bubble diagram with arbitrary external and internal particle masses has been calculated by \cite{Marolf:2010zp}. We briefly summarize the relevant results here. The two-point function from the bubble diagram is
\be
\expval{\sigma(x)\sigma(y)}_{\text{bubble}} = H^2\frac{\Gamma\left(\frac{3}{2}\right)}{2\cpi^{\frac{5}{2}}}\sum_{L=0}^{\infty}\frac{\left(L+\frac{3}{2}\right)g^2 \rho_{\sigma\chi}(L)}{\left(m_{\sigma}^2/H^2 + L(L+3)\right)^2}C_L^{\frac32}(Z_{xy}),
\ee
where $\rho_{\sigma\chi}(L)$ is a spectral function that decomposes the product of two propagators of unequal mass into a sum over Gegenbauer functions
\be
\Delta^{\sigma}(x,y)\Delta^{\chi}(x,y)\equiv H^4 \frac{\Gamma\left(\frac{3}{2}\right)}{2\cpi^{\frac{5}{2}}}\sum_{L=0}^{\infty}\left(L+\frac{3}{2}\right)\rho_{\sigma\chi}(L)C_L^{\frac32}(Z_{xy}).
\ee
This spectral function depends on $L$ and the masses of $\sigma$ and $\chi$. We will not go into the specific form of the $\rho$, which consists of $_7F_6$ hypergeometric functions. Details can be found in Appendix B of~\cite{Marolf:2010zp}. Given the expression for the bubble diagram, the half-sunset diagram is simply
\be
\begin{split}
\expval{\sigma(x)\sigma(y)}_{\text{half-sunset}} &=\expval{\chi(x)\chi(y)}_{\text{zero mode}}\times\expval{\sigma(x)\sigma(y)}_{\text{bubble}}\\
&= \frac{3}{8\cpi^2}\frac{H^2}{m_{\chi}^2}\times  H^2\frac{\Gamma\left(\frac{3}{2}\right)}{2\cpi^{\frac{5}{2}}}\sum_{L=0}^{\infty}\frac{\left(L+\frac{3}{2}\right)g^2 \rho_{\sigma\chi}(L)}{\left(m_{\sigma}^2/H^2 + L(L+3)\right)^2}C_L^{\frac32}(Z_{xy})\\
&\equiv H^2\frac{\Gamma\left(\frac{3}{2}\right)}{2\cpi^{\frac{5}{2}}}\sum_{L=0}^{\infty}\frac{\left(L+\frac{3}{2}\right)\Pi(L)_{\sigma\chi}}{\left(m_{\sigma}^2/H^2 + L(L+3)\right)^2}C_L^{\frac32}(Z_{xy}),
\end{split}
\ee
where on the last line we have introduced the combination
\be
\Pi_{\sigma\chi}(L)\equiv \frac{3}{8\cpi^2}\frac{H^2}{m_{\chi}^2}g^2 \rho_{\sigma\chi}(L)
\ee
for notational simplicity. The 1PI sum of half-sunset diagrams can be calculated just like we did for the zero mode seagull diagram:
\be
\begin{split}
&\begin{tikzpicture}[line width=1.5pt]
\draw (-2.5, 0)--(-0.5, 0);
\node at (-0.25, 0) {$+$};
\draw (0, 0)--(2, 0);
\draw[blue] (0.5,0) arc (180:0:0.5);
\draw[blue, dashed] (0.5,0) arc (180:360:0.5);
\node at (2.25, 0) {$+$};
\draw (2.5, 0)--(6, 0);
\draw[blue] (3,0) arc (180:0:0.5);
\draw[blue, dashed] (3,0) arc (180:360:0.5);
\draw[blue] (4.5,0) arc (180:0:0.5);
\draw[blue, dashed] (4.5,0) arc (180:360:0.5);
\node at (6.6, 0) {$+\cdots$};
\end{tikzpicture}\\
\equiv&\expval{\sigma(x)\sigma(y)}_{\text{1PI, half-sunset}}\\
=&H^2\frac{\Gamma\left(\frac{3}{2}\right)}{2\cpi^{\frac{5}{2}}}\sum_{L=0}^{\infty}\frac{\left(L+\frac{3}{2}\right)}{m_{\sigma}^2/H^2 + L(L+3)-\Pi_{\sigma\chi}(L)}C_L^{\frac32}(Z_{xy}).
\end{split}
\ee

Now, because $\Pi(L)_{\sigma\chi}$ is a complicated function of $L$, it is not so clear what the physical effect of the 1PI sum of half-sunset diagrams is from the above form. However, \cite{Marolf:2010zp} has shown that the large $Z_{xy}$ behavior of the two-point function is determined by the two $L$ poles of
\be
\frac{1}{m_{\sigma}^2/H^2+L(L+3)-\Pi_{\sigma\chi}(L)}.
\ee
In particular, at large $Z_{xy}$, the $\sigma$ two-point function behaves as $Z_{xy}^{L_{1,2}}$, where $L_{1,2}$ are the two poles. We care precisely about the large $Z_{xy}$ behavior of the $\sigma$ two-point function because it corresponds to the squeezed limit behavior of the inflaton three-point function that $\sigma$ mediates. The $L$ poles of the above function can only be solved for perturbatively, expanding in small $\Pi/m_\sigma^2$, which gives
\be
L_{\pm} = -\frac{3}{2}+\frac{\Im \Pi_{\sigma\chi}(L=-\frac{3}{2}+\iu \nu_{\sigma})}{\nu_{\sigma}} \pm \iu \left[\nu_{\sigma}-\frac{\Re \Pi_{\sigma\chi}\left(L=-\frac{3}{2}+\iu \nu_{\sigma}\right)}{2\nu_{\sigma}}\right],\quad \nu_{\sigma}\equiv \sqrt{m_{\sigma}^2-\frac{9}{4}}.
\ee
Coming back to our original goal, which is to find a distinctive modification of the $\sigma$ two-point function due to the presence of $\chi$, recall that the \textit{free} $\sigma$ two-point function has a large $Z_{xy}$ behavior
\be
\expval{\sigma(x)\sigma(y)}_{\text{free}}\xrightarrow{\text{large } Z_{xy}}Z_{xy}^{L_{\pm,\text{free}}}, \quad L_{\pm,\text{free}} = -\frac{3}{2}\pm \iu \nu_{\sigma}.
\ee
Several comments are in order about the difference between $L_{\pm}$ and $L_{\pm,\text{free}}$: first of all, we see that the real part of $\Pi\left(L=-\frac{3}{2}+\iu \nu_{\sigma}\right)$ is effectively contributing a correction to the mass of $\sigma$. And it turns out that this contribution is divergent as the dimension of the Euclidean de Sitter spacetime approaches four. As we have discussed during our calculation of the zero-mode seagull diagram, the counterterm $\delta_{m_{\sigma}}$ must be chosen at $\mathcal{O}(g^2)$ to cancel out this correction and maintain $m_{\sigma}$ as the physically observed mass. In other words, the half-sunset diagram turns out to contain a local-type correction as well, which is removed by our renormalization condition just like all other local-type corrections. 

\begin{figure}[!t]
    \centering
    \includegraphics[width=0.49\textwidth]{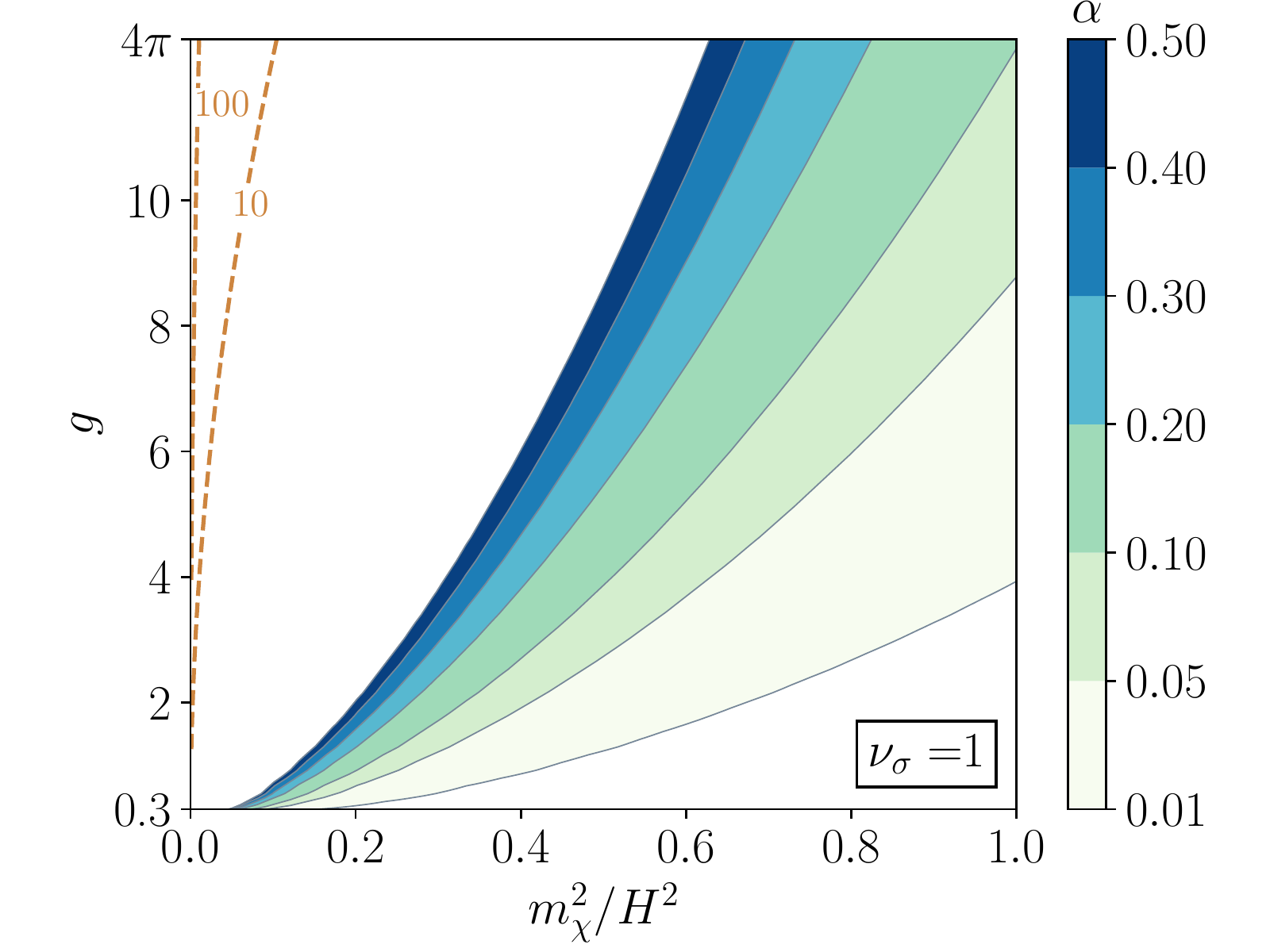}
    \includegraphics[width=0.49\textwidth]{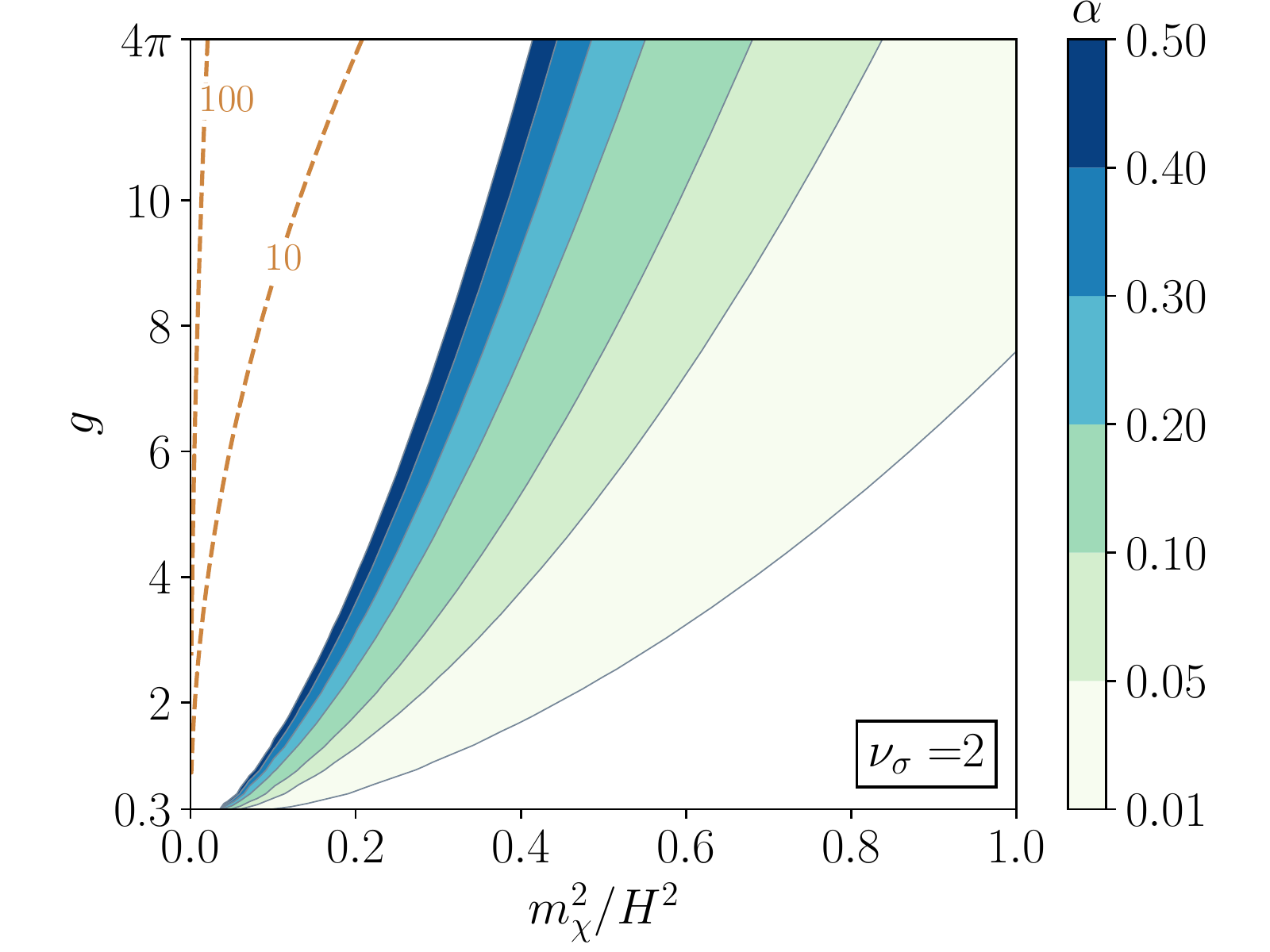}
    \includegraphics[width=0.49\textwidth]{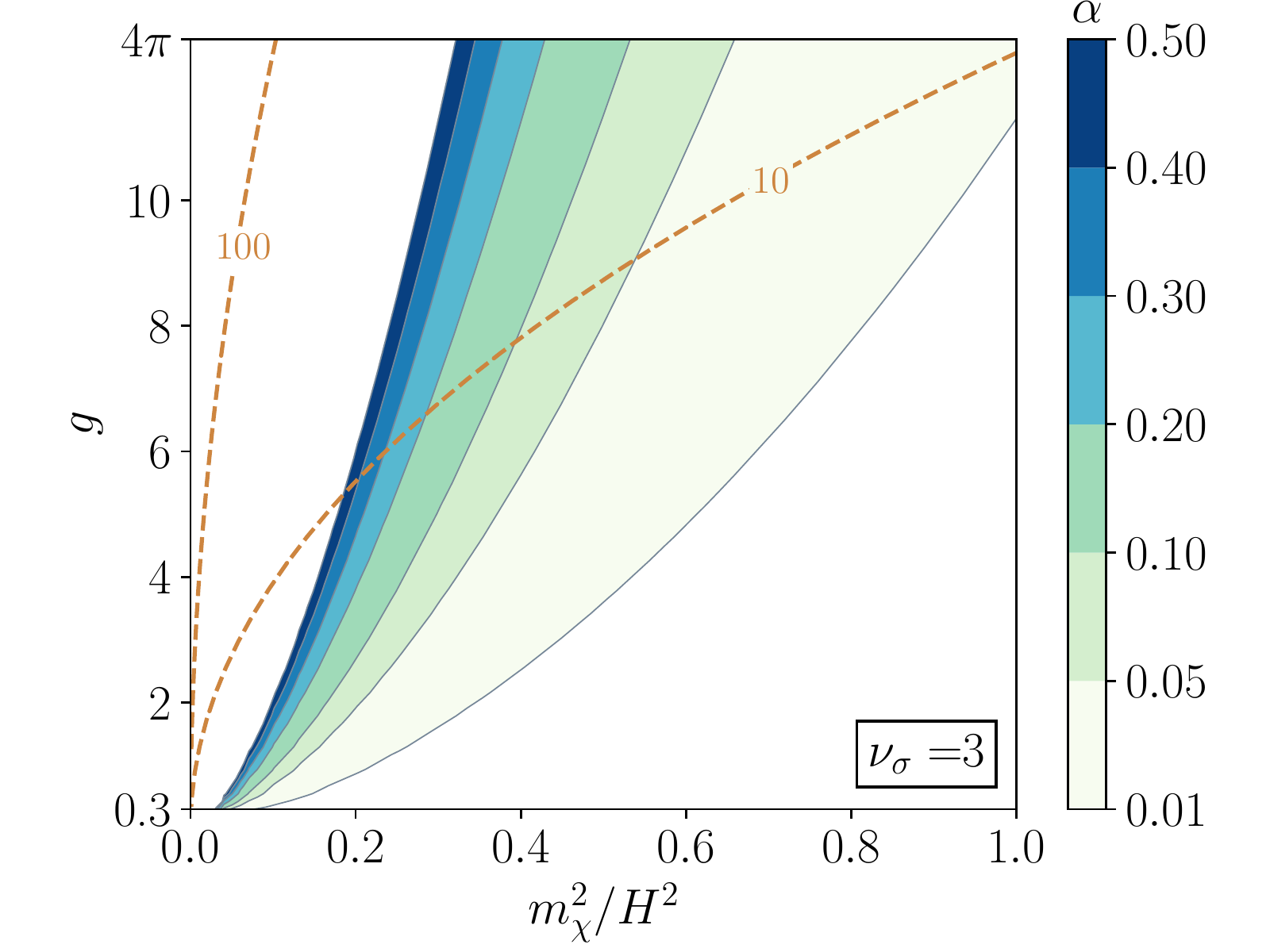}
    \includegraphics[width=0.49\textwidth]{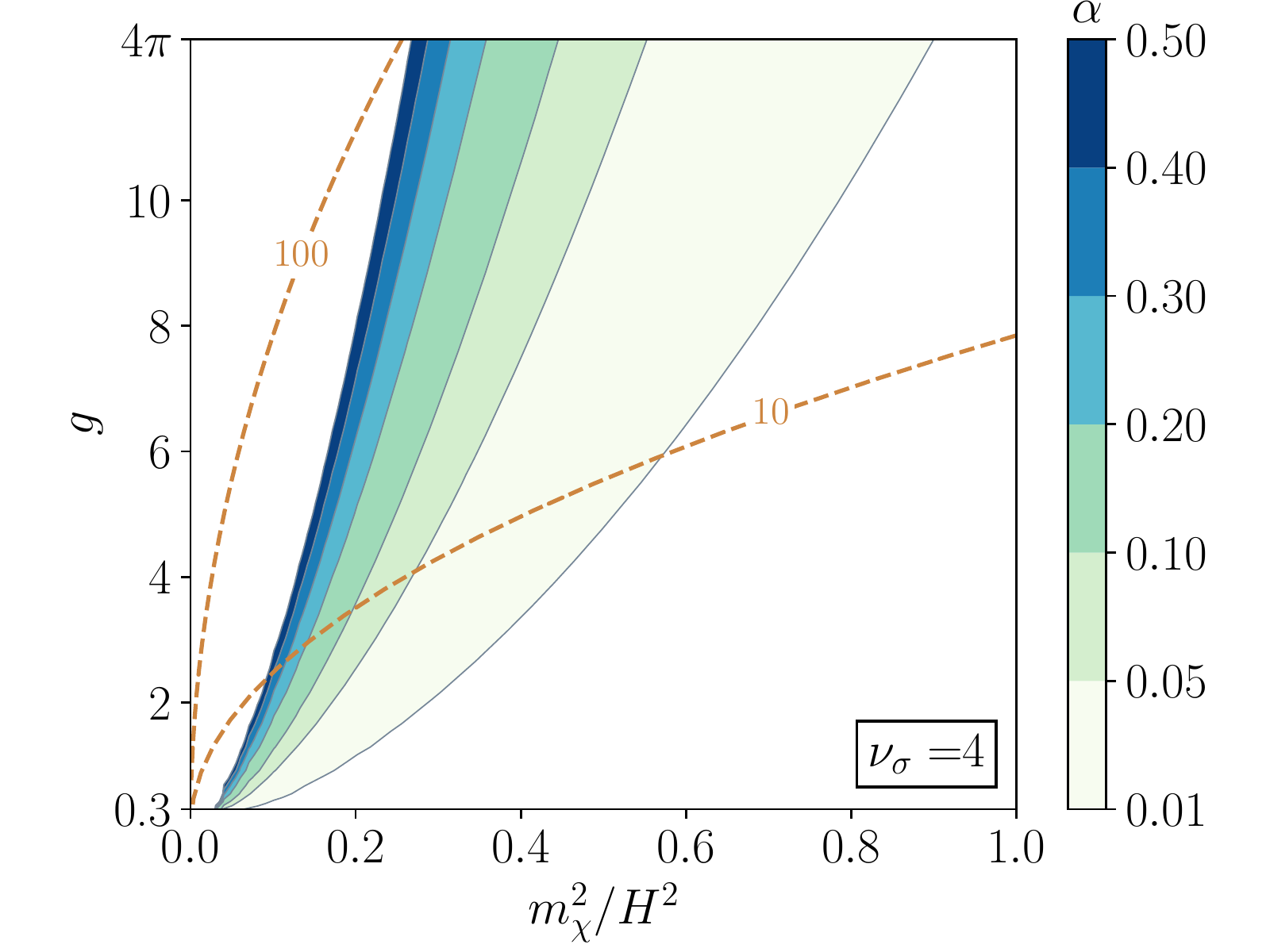}
    \caption{Dependence of the signal strength $\alpha$ (defined in \Eq{eq:alphadef}, which alters the signal scaling as in \Eq{eq:signalscaling}) on $m_{\chi}^2$ and $g$ for fixed $m_{\sigma}$ with $\nu_{\sigma} = 1$, 2, 3, 4. The range of $\alpha$ that is plotted is approximately the range that we expect to be observable in future 21 cm observations, as discussed in \Sec{sec:observability}. The dashed brown contours are where the calculable 1-loop correction~\Eq{eq:finetuning} to $\delta m_\chi^2$ is 10 and 100 times the physical value, respectively. These contours are intended to give a very rough guide to which regions of parameter space are necessarily fine tuned.}
    \label{fig:signal}
\end{figure}

Clearly, what is more interesting is the effect of the imaginary part of $\Pi$, which cannot be mimicked by a changed in $m_{\sigma}$. Within the range of $g$, $m_{\chi}^2$ and $m_{\sigma}^2$ values that we have considered (shown in \Fig{fig:signal}), $\Pi\left(L=-\frac{3}{2}+\iu \nu_{\sigma}\right)$ turns out to always have a negative imaginary part, 
\begin{equation} \label{eq:alphadef} 
\Im \Pi\left(L=-\frac{3}{2}+\iu \nu_{\sigma}\right)\equiv -\alpha.
\end{equation} 
This means the $\sigma$ two-point function is decaying faster with distance when it interacts with $\chi$. In other words, the $\sigma$ field values at different points in space are less correlated with each other. This loss of correlation can be understood as the following: because $\chi$ is light compared to the Hubble scale, its field value has $\mathcal{O}(H)$ fluctuations in space at super-Hubble scales. To calculate the half-sunset diagram, the two interaction points are individually integrated over all of space. When the two interaction points lie at a super-Hubble distance apart, they will see different $\chi$ field values, leading to different corrections to $m_{\sigma}$. This mismatch in correction to $m_{\sigma}$, once integrated over all space, manifests as a loss of correlation in $\sigma$.

To have the complete form of the inflaton bispectrum mediated by the $\chi$-loop corrected $\sigma$ two-point function, we would need to compute the diagram in \Eq{eq:bispec} using the modified $\sigma$ two-point function. However, the momentum dependence of the inflaton bispectrum in the squeezed limit can be extracted without such a calculation. In the scenario where $\sigma$ has no interaction with $\chi$, the squeezed limit of the inflaton bispectrum mediated by $\sigma$ is the standard quantum clock signal~\cite{Chen:2015lza}: when the subhorizon inflaton momentum matches the massive $\sigma$'s oscillation frequency, there is a resonance and the oscillation frequency is imprinted on the inflaton bispectrum. In our case, the inflaton bispectrum mediated by the loop-corrected $\sigma$ sees the same massive oscillation frequency, except now $\expval{\sigma(x)\sigma(y)}$ is smaller at larger distances. This means that the inflaton bispectrum has smaller amplitude when the inflaton mode resonates with $\sigma$ mode further in the past, i.e., at larger squeezedness. In other words, the loss of correlation in the $\sigma$ two-point function is directly translated to a loss of correlation in the inflaton bispectrum that $\sigma$ mediates,
\be \label{eq:signalscaling}
S_{\text{signal}}(k_1=k_2 \gg k_3) \propto  \left(\frac{k_3}{k_1}\right)^{\frac12+\iu\nu_{\sigma}+\alpha}.
\ee
Notice the flip of sign in front of $\alpha$, which comes from the fact that large distance in real space corresponds to smaller $k_3/k_1$ ratio. We were only able to extract the large distance (or small $k_3/k_1$) behavior of the 1PI sum of half-sunset diagrams that modifies the $\sigma$ two-point function. But this turns out to be advantageous for observational purposes, because the loss of coherence is more prominent at large distances.

The dependence of $\alpha$ on $m_{\sigma}$, $m_{\chi}$, and $g$ is shown in \Fig{fig:signal}. For each panel, the physical $\sigma$ mass is fixed such that $\nu_{\sigma} = 1$, 2, 3, 4. For fixed $\nu_{\sigma}$, there is a degeneracy in the $m_{\chi}^2-g$ plane when the two variables both decrease. When $m_{\chi}^2$ is smaller, the $\chi$ fluctuation in space is larger and leads to a stronger loss of correlation in $\sigma$, but this is compensated by a smaller coupling that suppresses the signal. For fixed $m_{\chi}^2$ and $g$, greater $m_{\sigma}^2$ leads to a smaller signal, since the greater the $\sigma$ mass, the less important is the effect of the space-dependent mass correction $\sigma$ receives from its interaction with $\chi$. The choice of range of $\alpha$ shown in \Fig{fig:signal}, $\alpha\in [0.01, 1/2]$, is based on a preliminary Fisher forecast estimate of which values can provide a detectable signal in future 21 cm observations, as described in the next section.

Readers familiar with the literature might recall that a similar loss of correlation has been observed when $\sigma$ has a self-interaction, such as $g_3\sigma^3$ \cite{Jatkar:2011ju, Boyanovsky:2012qs, Arkani-Hamed:2015bza, Krotov:2010ma}. In this context, the loss of correlation in the $\sigma$ two-point function has been interpreted as a $\sigma\to 2\sigma$ ``decay'' process, which is kinematically forbidden in Minkowski space, but possible in de Sitter due to the thermal-like environment. Our missing scalar signal can also be interpreted as a contribution to the decay width of $\sigma$. The analogous ``decay'' process is $\sigma \to \sigma + 2\chi$. However, this does not mean that the effect of a $\sigma$ self-interaction is indistinguishable from the missing scalar signal, because the self-interaction contribution to $\alpha$ is much suppressed compared to the missing scalar contribution. There are two sources of this suppression. First of all, there is no enhancement coming from the zero mode of a light scalar. Secondly, the $\sigma\to 2\sigma$ ``decay'' process is made possible only due to the thermal-like environment of de Sitter, meaning that the decay width is exponentially suppressed when the decay product $\sigma$ is heavier than Hubble. For the missing scalar signal, there is no similar suppression since the analogous decay process is $\sigma \to \sigma + 2\chi$, where no extra $\sigma$ is being produced from the vacuum. Using results from \cite{Jatkar:2011ju}, even with $g_3 = m_{\sigma}$, which is the maximum value before the theory becomes strongly coupled, the contribution to $\alpha$ due to a $g_3\sigma^3$ self-interaction is found to be less than $0.001$ for $\nu_{\sigma} = 1$ and 2. For heavier $\sigma$, the exponential suppression is going to be even stronger. Therefore, within the range of $\alpha\in [0.01, 1/2]$ that is potentially detectable in future 21 cm observations, $\alpha$ is unambiguous evidence for the existence of a light scalar $\chi$ interacting with $\sigma$.

When $m_\chi^2 \ll m_\sigma^2$, the counterterms must be adjusted very carefully to preserve the mass hierarchy between the two scalars. This is an example of the usual fine-tuning problem associated with interacting scalar fields. In principle, $m_\chi^2$ is subject to corrections from arbitrarily heavy mass scales in the theory, but the problem is at least as severe as the calculable contributions associated with $m_\sigma^2$ itself. The flat-space Coleman-Weinberg potential calculated in dimensional regularization with renormalization scale set equal to the Hubble scale $H$ contains a correction to the mass term of $\chi$ given by
\begin{equation} \label{eq:finetuning}
\delta m_\chi^2 = \frac{1}{16\pi^2} g^2 m_\sigma^2 \left(\log \frac{m_\sigma^2}{H^2} - \frac{3}{2}\right).
\end{equation}
To provide a sense of the minimum level of fine tuning of the scalar potential associated with our signal, we have plotted contours of $|\delta m_\chi^2| = 10 m_\chi^2$ and $|\delta m_\chi^2| = 100 m_\chi^2$ in \Fig{fig:signal}. Of course, if the UV cutoff of the theory is far above $m_\sigma$, the true fine tuning could be much worse. Although the light scalar $\chi$ must be tuned to be light, one could generalize our results to consider interactions of the form $A_\mu A^\mu |\sigma|^2$ associated with charged scalars $\sigma$ obtaining a spatially varying mass from fluctuations of the gauge field $A_\mu$. In such a case, there would be no fine tuning, because $A_\mu$ is naturally light. It would be interesting to see if similar signal strengths could be obtained in such a modified scenario. We leave such considerations for future work. Another possible scenario to circumvent the fine-tuning problem is to impose a shift symmetry on $\chi$ which is only softly broken by the $\chi$ mass. This requires the coupling between $\sigma$ and $\chi$ to be in derivative form, such as $\frac{\sigma^2(\partial_{\mu}\chi)^2}{\Lambda_{\chi}^2}$.\footnote{We discuss here a $\sigma^2(\partial_{\mu}\chi)^2$ coupling instead of $\sigma(\partial_{\mu}\chi)^2$ given the $\mathbb{Z}_2:\, \sigma\rightarrow -\sigma$ symmetry in our original model.} We expect the derivative coupling to produce a similar loss of correlation effect. The main mechanism of our missing scalar signal is the spatial variation in the insertion that $\sigma$ sees, and indeed $\partial_{\mu}\chi$ has spatial variation in its value. But the contribution to $\alpha$ from such derivative coupling would be suppressed compared to that from a direct coupling because of the lack of $\chi$ zero-mode enhancement as well as the smaller coefficient from a higher-dimension operator.

\section{Preliminary Estimate of Observability}
\label{sec:observability}

We have seen that the interaction of $\sigma$ with a light field $\chi$ leads to an extra suppression of the non-Gaussanity by $(k_3/k_1)^\alpha$ (where $k_3 \ll k_1$ and $\alpha > 0$), as in \Eq{eq:signalscaling}. We would like to determine what range of $\alpha$ is potentially observable. In this section, we present a preliminary estimate that the range $\alpha\in[0.01, 0.5]$, as depicted in \Fig{fig:signal}, may be observable in future 21 cm observations. A full analysis could apply a Fisher forecast to assess how well future large scale structure or 21 cm observations could be used to infer the parameters of a signal template incorporating the inflaton bispectrum's dependence on the amplitude $f_{\text{NL}}$, the oscillation frequency $\nu_{\sigma}$, and the loss of correlation $\alpha$. We leave such a detailed study for future work, instead providing an approximate argument for the observability of $\alpha$ based on an existing analysis of the observability of the ``clock'' signal mediated by a heavy scalar interacting with the inflaton. We will summarize the method of the Fisher forecast that we are borrowing, what assumptions are necessary to obtain a sensitivity on $\alpha$, and the results.

The Fisher forecast analysis we are relying on comes from~\cite{Meerburg:2016zdz}, which forecasts the sensitivity of future 21cm experiments for the inflaton bispectrum mediated by a freely propagating scalar.\footnote{For related forecasts of the sensitivity of large-scale structure and 21 cm observations to inflationary features and clock signals, see, e.g.,~\cite{Chen:2016zuu, Chen:2016vvw, Ballardini:2016hpi, Xu:2016kwz, MoradinezhadDizgah:2018ssw}.} The paper studies the inflaton bispectrum mediated by both light ($m^2/H^2 < 9/4$) and heavy ($m^2/H^2>9/4$) scalars, but we will focus on the heavy scalar which is relevant for our signal. The inflaton three-point correlation function is parameterized as
\be
\left\langle\zeta^{3}\right\rangle \equiv(2 \pi)^{3} \delta^{(3)}\left(\mathbf{k}_{123}\right) \frac{A^{2}}{\left(k_{1} k_{2} k_{3}\right)^{2}} S\left(k_{1}, k_{2}, k_{3}\right) \equiv B\left(k_{1}, k_{2}, k_{3}\right)(2 \pi)^{3} \delta_{D}\left(\mathbf{k}_{123}\right),
\ee
where $A = 2\pi^2 A_{s}$, $S(k_1, k_2, k_3)$ is the shape function, and $B$ is the bispectrum. The shape function template for the inflaton bispectrum mediated by a heavy scalar $\sigma$ is
\be
S^{\text{clock}}(k_1,k_2,k_3) = f_{\text{NL}}\frac{3^{\frac72}}{2}A(\alpha_{123})(\alpha_{123})^{-\frac12}\sin\left(\nu_{\sigma}\ln(\frac{\alpha_{123}}{2})+\delta\right)+\text{2 perm},
\ee
where $\alpha_{123} = \frac{k_1+k_2}{k_3}$, $\delta$ is a calculable but model-dependent phase, and we retain our earlier notation $\nu_{\sigma} = \sqrt{m_{\sigma}^2/H^2-9/4}$ (which was denoted $\mu$ in~\cite{Meerburg:2016zdz}). In the rest of the analysis, the phase $\delta$ is assumed to be zero, and the forecast result only has $\mathcal{O}(1)$ dependence on the value of the phase. $A(\alpha_{123})$ is a window function meant to remove equilateral contributions, whose details will not be relevant for our purposes. 
We haven't said anything about the overall size $f_\text{NL}$ of our signal, which obviously depends on model details such as the coupling between $\si$ field and the inflaton. It is possible to embed our signal into known cosmological collider models of which the signal sizes were calculated. We don't expect that the presence of a nonzero $\al$ can significantly affect the signal size. See \cite{Wang:2019gbi} for a summary of models with visibly large $f_\text{NL}$. 

The Fisher matrix is defined to be
\be
F_{i j}^{b}=\sum_{z_{k}} \sum_{T, T^{\prime}} \frac{\partial B\left(T, z_{k}\right)}{\partial p_{i}}\left(C^{-1}\right)_{T T^{\prime}} \frac{\partial B\left(T^{\prime}, z_{k}\right)}{\partial p_{j}},
\ee
where $p_i$ are the parameters that the bispectrum depends on, $z_k$ are redshifts, $T$ is the sum over triangles in momentum space
\be
\sum_{T} \equiv \sum_{k_{1}=k_{\min }}^{k_{\max }} \sum_{k_{2}=k_{1}}^{k_{\max }} \sum_{k_{3}=\max \left(k_{\min }, k_{2}-k_{1}\right)}^{k_{2}},
\ee
and the Gaussian covariance matrix between triangle configurations is~\cite{Scoccimarro:2003wn, Baldauf:2016sjb}
\be
C_{T T^{\prime}}=\frac{(2 \pi)^{3}}{V_{i}} \frac{\pi s_{123}}{d k_{1} d k_{2} d k_{3}} \frac{P\left(k_{1}\right) P\left(k_{2}\right) P\left(k_{3}\right)}{k_{1} k_{2} k_{3}} \delta_{T T^{\prime}}
\ee
with symmetry factor $s_{123}$ ($  = 6,~2,~1$ for equilateral, isosceles and general triangles).

The parameters $p_i$ labeling rows of the Fisher matrix in~\cite{Meerburg:2016zdz} are the amplitude $f_{\text{NL}}$ of the signal, the oscillation scale $\nu_\sigma$, and additional parameters characterizing secondary non-Gaussianities. Using the $S^{\text{clock}}$ template, \cite{Meerburg:2016zdz} finds that for $f_{\text{NL}} = 1$, 21 cm experiments can achieve a sensitivity of $\Delta\nu_{\sigma}\approx 0.01$ for $\nu_{\sigma} = 0.7$, 1.0, and 3.0. With smaller $f_{\text{NL}}$, the sensitivity to $\nu_{\sigma}$ deteriorates with $\Delta\nu_{\sigma} \sim f_{\text{NL}}^{-1}$.

With a few assumptions, the above results can be translated to a sensitivity on $\alpha$, even though the signal template used contained no dependence on $\alpha$. The inflaton bispectrum from our missing scalar signal can be described by the template
\be
S^{\text{MS}}(k_1, k_2, k_3) =  f_{\text{NL}}\frac{3^{\frac72}}{2}A(\alpha_{123})(\alpha_{123})^{-\frac12-\alpha}\sin\left(\nu_{\sigma}\ln(\frac{\alpha_{123}}{2})+\delta\right)+\text{2 perm},
\ee
which is exactly the same as the previous $S^{\text{clock}}$ template except the additional $-\alpha$ power on $\alpha_{123}$.

Comparing our missing scalar signal template and the clock template, the derivative with respect to $f_{\text{NL}}$ is the same except the difference in overall momentum scaling due to $\alpha$. If we focus on the observability of small values of $\alpha$, it is reasonable to assume that the additional $\alpha$ power makes no significant difference in the forecast of sensitivity to $f_{\text{NL}}$. The same holds for the derivative with respect to $\nu_{\sigma}$. However, we now have an additional row and column of the Fisher matrix thanks to the additional parameter $\alpha$. 

The new entries in the Fisher matrix depend on the derivatives of the oscillation template with respect to $\alpha$. The existing analysis used the derivative of the oscillation template with respect to $\nu_{\sigma}$,
\be
\pdv{S}{\nu_{\sigma}} = f_{\text{NL}}\frac{3^{\frac72}}{2}A\left(\frac{k_1+k_2}{k_3}\right)\left(\frac{k_1+k_2}{k_3}\right)^{-\frac12}\cos\left(\nu_{\sigma}\log(\frac{k_1+k_2}{k_3})+\delta\right)\log(\frac{k_1+k_2}{k_3})+\text{2 perm.}
\ee
The derivative of the missing scalar template with respect to $\alpha$ is
\begin{align}
\pdv{S_{\text{MS}}}{\alpha} &= f_{\text{NL}}\frac{3^{\frac72}}{2}A\left(\frac{k_1+k_2}{k_3}\right)\log \left(\frac{k_1+k_2}{k_3}\right)\!\!\left(\frac{k_1+k_2}{k_3}\right)^{-\frac12+\alpha}\!\!\sin\left(\nu_{\sigma}\log(\frac{k_1+k_2}{k_3})+\delta\right)\!+\text{2 perm.} \nonumber \\
&= \pdv{S}{\nu_{\sigma}}\left(\delta\rightarrow \delta+\frac{\pi}{2}\right) + \mathcal{O}(\alpha).
\end{align}
These two derivatives are only different by a shift in the phase $\delta$. Assuming that the forecast result on $\nu_{\sigma}$ is independent of $\delta$, we can directly translate the projected absolute uncertainty on $\nu_{\sigma}$ to a projected absolute uncertainty on $\alpha$. We conclude that 21 cm experiments can measure $\alpha$ up to $\approx 0.01$ accuracy when $f_{\text{NL}} = 1$, leading to an observational lower bound of $\alpha > 0.01$.

The strong similarity between the expressions for $\partial S/\partial \nu_\sigma$ and $\partial S_\mathrm{MS}/\partial \alpha$ does not imply that there is a degeneracy between $\nu_\sigma$ and $\alpha$ in the fit. It may be useful to explain this more explicitly. Focusing on the 2-by-2 block of the Fisher matrix involving $\nu_\sigma$ and $\alpha$, it will schematically have the form
\begin{equation}
\begin{pmatrix}
\sum_T f(\{k\}_T) \cos^2(g(\{k\}_T)) & \sum_T f(\{k\}_T) \cos(g(\{k\}_T)) \sin(g(\{k\}_T)) \\
\sum_T f(\{k\}_T) \cos(g(\{k\}_T)) \sin(g(\{k\}_T)) & \sum_T f(\{k\}_T) \sin^2(g(\{k\}_T)) 
\end{pmatrix},
\end{equation}
where $\{k\}_T$ stands for the collection of momenta associated with a given triangle, and $f()$ and $g()$ are shorthand for the functions appearing in the prefactor and the argument of the (co)sine. Because the arguments of the (co)sines are different for different triangles, we expect that the off-diagonal terms of the form $\cos \times \sin$ will tend to average toward zero when sampling many triangles, whereas the diagonal $\cos^2$ and $\sin^2$ terms will have nonvanishing averages. As a result, we expect that $\nu_\sigma$ and $\alpha$ will be independently constrained, with similar uncertainties and without large covariance.
 
It is worth emphasizing here that the translation of the sensitivity on $\nu_{\sigma}$ from~\cite{Meerburg:2016zdz} to the sensitivity on $\alpha$ is only a leading order estimate. When $\alpha$ is larger, in particular when it is greater than $1/2$, the inflaton bispectrum decays relative to the equilateral shape. This does not necessarily mean that the signal is unobservable, but a separate Fisher forecast analysis is needed to take into account the dominating equilateral shape background. For simplicity, we set $\alpha < 1/2$ as the observable upper bound.

\section{Conclusions}
\label{sec:conclusions}

In the search for a better understanding of fundamental physics beyond the Standard Model, we have a unique opportunity to exploit the natural experiments carried out by the early universe, which accessed energy scales well above those we can probe with terrestrial experiments. It is up to us to read out the data from these experiments from the sky, and use it to its fullest potential. This requires a systematic theoretical exploration of the non-Gaussianities produced by the full range of possibilities for high-energy particle physics.

Oscillating signals in primordial non-Gaussianities can offer powerful probes of new physics, indicating the existence of very massive particles beyond the reach of colliders~\cite{Chen:2009we, Chen:2009zp, Arkani-Hamed:2015bza} or helping to distinguish the physics of inflation from alternatives~\cite{Chen:2011zf, Chen:2014cwa}. During inflation, fields with mass $m/H > 3/2$ can generate such signals. Scalar fields below this mass do not lead to the telltale oscillations. In this work, we have shown that they nonetheless can influence the oscillating signals produced by heavy scalars with which they interact. The characteristic signal is a suppression of the amplitude of the oscillatory non-Gaussianity in the squeezed limit by a factor $(k_3/k_1)^\alpha$, where $\alpha \propto g^2$ parameterizes the suppression of correlation in the heavy field by fluctuations in the value of the light field.

The minimal version of our scenario is potentially fine tuned, when there is a hierarchy in scalar mass $m_\chi^2 \ll m_\sigma^2$. An interesting follow-up would be to explore whether fluctuations in a gauge field $A_\mu$ can produce a similar effect in the non-Gaussianities mediated by charged scalars, perhaps evading the tuning of the all-scalar scenario.  Conversely, as noted in the introduction, we could consider a light scalar $\chi$ interacting with a higher-spin field $\sigma$, which could change the characteristics of the non-Gaussanity that $\chi$ modulates.

We have provided a preliminary estimate of the range of signal strengths that are observable, by noting that the form of the Fisher matrix entries determining the uncertainty in measurements of $\alpha$ is very similar to that of those determining the uncertainty in measurements of the oscillation rate $\nu_\sigma$. The observability of the latter using 21 cm data was previously examined in~\cite{Meerburg:2016zdz}. While this argument provides a first estimate of the smallest accessible values of $\alpha$ in data, at larger $\alpha$ the suppression of the oscillatory signal will make its reconstruction more difficult. This motivates a more detailed study of the full Fisher forecast for our model in the future.

\section*{Acknowledgments}

We thank Xingang Chen for useful comments on a draft of this paper. The work of QL and MR is supported in part by the NASA Grant 80NSSC20K0506 and the DOE Grant DE-SC0013607. ZZX is supported by Tsinghua University Initiative Scientific Research Program. This work was performed in part at the Aspen Center for Physics, which is supported by the National Science Foundation grant PHY-1607611.

\bibliographystyle{utphys}
\bibliography{ref}
\end{document}